%% file: Hohle_et_al_Lines.tex
\documentclass[useAMS,usenatbib]{mn2e}
\input definitions
\usepackage{graphicx}  
\usepackage{longtable}                
\usepackage{subfig}
\usepackage{caption}
\usepackage{everysel}
\usepackage{keyval}
\usepackage{ragged2e}
\usepackage{rotating}
\usepackage[draft,pdftex,pdfpagemode={UseOutlines},bookmarks,bookmarksopen,colorlinks,linkcolor={blue},citecolor={green},urlcolor={red}]{hyperref}
\bibpunct{(}{)}{;}{a}{}{,}

\title[Absorption features in the X-ray spectra of isolated Neutron Stars]{Narrow absorption features in the co-added \xmm{} RGS spectra of isolated Neutron Stars.}
\author[M. M. Hohle et al.]{M. M. Hohle$^{1}$\thanks{mhohle@astro.uni-jena.de}, F. Haberl$^{2}$\thanks{fwh@mpe.mpg.de} and J. Vink$^{3}$\thanks{j.vink@astro.uu.nl}, C.P. de Vries$^{4}$\thanks{C.P.de.Vries@sron.nl}, R. Neuh\"auser$^{1}$\thanks{rne@astro.uni-jena.de}\\
$^{1}$Astrophysikalisches Institut und Universit\"ats-Sternwarte Jena, Schillerg\"asschen 2-3, 07745 Jena, Germany\\
$^{2}$Max-Planck-Institut f\"ur extraterrestrische Physik, Giessenbachstra{\ss}e, 85741 Garching, Germany\\
$^{3}$University Utrecht, PO Box 80000, 3508 TA Utrecht, The Netherlands\\
$^{4}$ SRON, Netherlands Institute of Space Research, Sorbonnelaan 2, 3584 CA, Utrecht, The Netherlands}
\begin{document}

\bibliographystyle{mn2e}

\date{Accepted 201X Month XX. Received 201X Month XX; in original form 201X Month XX}

\pagerange{\pageref{firstpage}--\pageref{lastpage}} \pubyear{2002}

\maketitle

\label{firstpage}

\begin{abstract}
We co-added the available \xmm{} {\it RGS} spectra for each of the isolated X-ray pulsars \rxj{}, \rxd{} (\rbs), \rxe{} and \rxf{} (four members of the ``Magnificent Seven") and the ``Three Musketeers" Geminga, PSR\,B0656+14 and PSR\,B1055-52. We confirm the detection of a narrow absorption feature at $0.57~\mathrm{keV}$ in the co-added {\it RGS} spectra of \rxj{} and \rxe{} (including most recent observations). In addition we found similar absorption features in the spectra of \rxd{} (at $0.53~\mathrm{keV}$) and maybe PSR\,B1055-52 (at $0.56~\mathrm{keV}$). The absorption feature in the spectra of \rxd{} is broader than the feature e.g. in \rxj{}. The narrow absorption features are detected with 2$\sigma$ to 5.5$\sigma$ significance.\\
Although very bright and frequently observed, there are no absorption features visible in the spectra of \rxf{} and PSR\,B0656+14, while the co-added \xmm{} {\it RGS} spectrum of Geminga has not enough counts to detect such a feature.\\ 
We discuss a possible origin of these absorption features as lines caused by the presence of highly ionised oxygen (in particular OVII and/or OVI at $0.57~\mathrm{keV}$) in the interstellar medium and absorption in the neutron star atmosphere, namely the absorption features at $0.57~\mathrm{keV}$ as gravitational redshifted ($g_{r}$=1.17) OVIII.\\

\end{abstract}


\section{Introduction}

Neutron stars (NSs) give us the opportunity to study matter under extreme conditions, such as high densities ($\rho\approx 10^{15}~\mathrm{g/cm^{3}}$), strong magnetic fields ($\mathrm{B\approx10^{13}~G}$) and high temperatures ($\mathrm{T\approx10^{6}~K}$). Such conditions cannot be generated in laboratories and theoretical predictions need observational input to understand satisfactorily the behaviour of matter under these conditions, in particular for the interior of NSs. One main goal of NS astrophysics is to constrain the Equation of State (EoS) at supra-nuclear densities that would give us fundamental insights into various fields of physics, therefore suitable targets have to be found.\\
Most of the $\approx2000$ known NSs \citep{2005yCat.7245....0M} are radio pulsars, among them nine double NSs were found so far (see \citealt{2006JPhG...32S.259S} for a review) that enable us to determine their masses with high accuracy. However, it is not possible to infer the radius or compactness of these objects. In principle, the radius could be measured from the luminosity of cooling (hence, blackbody radiation) NSs, if the distance and temperature are known, using the modified Stefan-Boltzman law. If spectral lines could be detected as well, they deliver an estimate for the compactness of the particular NS from measuring the gravitational redshift $g_{r}$\footnote{Here, $g_{r}=1/\sqrt{1-r_{s}/R_{NS}}$, where $r_{s}$ denotes the Schwarzschild radius.}.\\
On the other hand, spectral features can be caused by the interstellar medium (ISM): The Sun is located in a 1MK hot cavity of $\mathrm{\sim300~pc}$ radius, the so-called Local Bubble \citep{2003A&A...411..447L} that was formed by multiple supernovae in the last few Myrs~\citep{2006A&A...452L...1B,2009SSRv..143..263B}. The Local Bubble (LB) is filled with a thin hot plasma including highly ionised mid Z-elements (in particular He-like oxygen). The exact temperature of the LB is currently under debate~\citep{2009AIPC.1156....3K,2009SSRv..143..231S}, but this environment likely causes absorption features in the soft X-rays. Thus, narrow absorption features in the spectra of isolated X-ray sources can be caused by the ISM and/or by intrinsic processes.\\
We selected those isolated NSs for our work that are bright in the X-rays, have simple spectral properties and that are observed deep enough with \xmm{} {\it RGS}, hence four members of the ``Magnificent Seven" (\rxj{}, \rxd{}\footnote{\rbs}, \rxe{} and \rxf{}) and all members of the ``Three Musketeers" (Geminga, PSR\,B0656+14 and PSR\,B1055-52). We co-added the high resolution {\it RGS} spectra of these targets and searched for narrow, maybe redshifted, absorption lines.

\subsection{The ``Magnificent Seven"}

During the ROSAT All Sky Survey, seven soft X-ray sources, the so called ``Magnificent Seven" (hereafter M7), were discovered that are suitable targets for such investigations. The M7 exhibit soft X-ray spectra that are best modelled with blackbody radiation ($\mathrm{kT_{eff}=40-110~eV}$) with broad absorption features in some cases. These absorption features are interpreted as proton-cyclotron resonances or atomic transitions of bound species in a strong magnetic field at $\mathrm{B\approx10^{13}~G}$. No radio emission could be detected from the M7 (see e.g. \citealt{2009ApJ...702..692K}), but they exhibit X-ray pulsations with periods between $\mathrm{3~s}$ and $\mathrm{12~s}$. All M7 are isolated, i.e. they are not associated with a supernova remnant, do not belong to a binary system and no sub-stellar companions are found so far \citep{2009A&A...496..533P}. The kinematic and characteristic age estimates yield values of $\mathrm{0.3-3~Myrs}$ \citep{2007Ap&SS.308..181H,2010MNRAS.402.2369T}. The two brightest members, \rxf{} and \rxj{}, have known trigonometric parallaxes \citep[yielding distances of $\mathrm{\sim125~pc}$ and $\mathrm{\sim360~pc}$, respectively]{2010arXiv1008.1709W,2007ApJ...660.1428K}\footnote{For \rxj{}, \citet{EisenPhd} obtains a distance of $\mathrm{280^{+210}_{-85}~pc}$.}, which allows one to derive limits on the size of the emitting area from the analysis of the X-ray spectrum. For a review of the M7 we refer to \citet{2007Ap&SS.308..181H} and \citet{2009ApJ...705..798K}.\\

\subsection{The ``Three Musketeers"}

Like the M7, the ``Three Musketeers" (hereafter 3M) form an own sub-sample of NSs with similar properties \citep{1997A&A...326..682B}. Their soft X-ray spectra are best modelled with two blackbodies of different normalisations and a hard power-law tail dominating above 2~keV (see \citealt{2005ApJ...623.1051D} for a review). The magnetic fields of the 3M ($\mathrm{B\approx10^{12}~G}$) are a factor of ten weaker than those of the M7, but the 3M are younger by trend (a few $\mathrm{10^{5}~yrs}$). Geminga (discovered by \citealt{1983ApJ...272L...9B}) and PSR\,B0656+14 (discovered by \citealt{1989ApJ...345..451C}) have distances derived from trigonometric parallaxes \citep{1996ApJ...461L..91C,2003ApJ...593L..89B,2005nscf.confE..23W} ranging from 150~pc to 300~pc, while the distance of PSR\,B1055-52 (750~pc) is estimated from dispersion measurements \citep{2003MNRAS.342.1299K}. Geminga and PSR\,B1055-52 also exhibit pulsations in gamma-rays \citep{1993ApJ...413L..27F,1992Natur.357..306B}.

\section[]{\xmm{} observations}

All \xmm{} \textbf{R}eflection \textbf{G}rating \textbf{S}pectrometer ({\it RGS}, \citealt{2001A&A...365L...7D}) observations have been analysed and reprocessed with the standard \xmm\ {\bf S}cience {\bf A}nalysis {\bf S}ystem (SAS) version 10.0 using the {\it rgsproc} task. We checked the background of each observation individually and created {\bf G}ood {\bf T}ime {\bf I}ntervall (GTI) files that filter out those times with count rates above 0.25 or 0.1 cts/s (depending on the source and background behaviour) determined for CCD number nine of {\it RGS1} and ran {\it rgsproc} again for GTI filtering. The photons for the spectra were obtained from the default {\it RGS} extraction regions and we co-added the spectra using {\it rgscombine} (that also produces co-added background spectra and the corresponding response matrices) allowing to investigate the total average spectrum with high energy resolution. The {\it RGS2} detector does not provide spectral information in the required energy band (due to the failure of chip four) around 0.57~keV, thus could not be used.\\
The co-added spectra are fitted and investigated using {\it Xspec~12}. We explain the data handling and fitting and fit results of the individual sources in the following sections.


\subsection{\rxj{}}

\begin{figure}
\centering
\includegraphics*[viewport=95 1 605 435, width=0.425\textwidth]{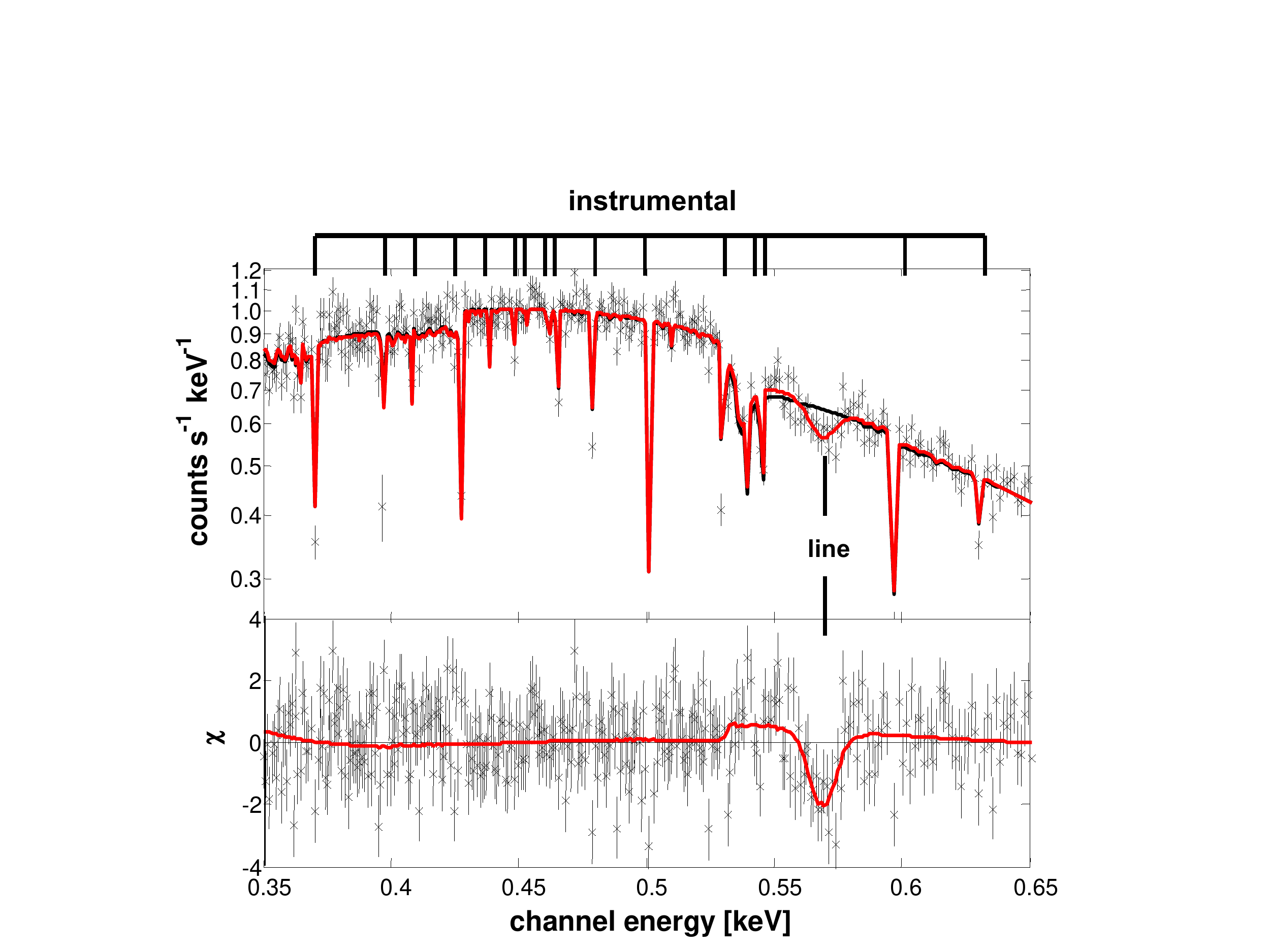}
\caption[]{The co-added {\it RGS1} (first order) spectrum of \rxj{} with a total exposure of $\mathrm{516~ks}$. The thin, solid black line represents the fit model, while the lower red line shows the fit model plus a narrow absorption feature seen at $\mathrm{568.6_{-1.9}^{+1.8}~eV}$ with $\mathrm{EW=-1.89_{-0.62}^{+0.56}~eV}$ (errors denote $90\%$ confidence level), see also \cite{2009A&A...497L...9H} for comparison. We also mark the locations of instrumental features exemplarily in this figure (not shown in the spectra of the other neutron stars).}
\label{RGS0720}%
\end{figure}

In contrast to the other M7, \rxj{} shows significant spectral and temporal variations on time scales of years that might be caused by a glitch (\citealt{2007ApJ...659L.149V}, see also \citealt{2005ApJ...628L..45K}) or free precession \citep{2009A&A...498..811H,2010A&A...521A..11H,2007Ap&SS.308..181H,2006A&A...451L..17H,2004A&A...415L..31D}. Due to this peculiarity, \rxj{} is one of the best investigated members of the M7. Since more than ten years \rxj{} is frequently observed with \xmm{} and {\it Chandra}.\\ 
After co-adding all available \xmm{} {\it RGS} spectra of \rxj{}, \citet{2009A&A...497L...9H} reported the detection of a narrow absorption feature at $\mathrm{0.57~keV}$. This feature was interpreted as the $K_{\alpha}$ absorption line of highly ionised oxygen: either OVI and/or OVII at rest, or OVIII (in this case $\mathrm{Ly_{\alpha}/K_{\alpha}}$) with a gravitational redshift of $g_{r}=1.16$, see \citet{2009A&A...497L...9H}. In \citet{2009A&A...497L...9H} it is argued that this absorption feature, if oxygen, may originate from the ambient medium of the NS, maybe from a circumstellar disk surrounding \rxj{}. The presence of such a disk may account partly for some of the peculiarities of \rxj{} \citep{2009A&A...497L...9H}.\\
Since the work of \citet{2009A&A...497L...9H}, we performed four further \xmm{} observations. We re-investigated the data used by \citet{2009A&A...497L...9H} together with the new \xmm{} observations, that adds up to 650~ks in total (533~ks after GTI filtering), see \autoref{RGS_0720_tab}.\\
\begin{table}
\centering
\caption[]{The \xmm{} {\it RGS1} observations of \rxj{} in chronological order. We list the exposure times and the net counts after the data passed the GTI filters.}
\label{RGS_0720_tab}

\begin{tabular}{ccrr}

\hline
Obs. ID & date & counts      & eff. exposure \\
	      &      & 0.35-1.0~keV & [ks]          \\
\hline
0124100101 & 2000 May 13 &  11480    &    42.36 \\
0132520301 & 2000 Nov 21 &   7878    &    30.24 \\
0156960201 & 2002 Nov 6  &   7518    &    29.61 \\
0156960401 & 2002 Nov 8  &   7796    &    31.22 \\
0158360201 & 2003 May 2  &  15507    &    61.56 \\
0161960201 & 2003 Oct 27 &  13513    &    44.67 \\
0164560501 & 2004 May 22 &   8007    &    26.97 \\
0300520201 & 2005 Apr 28 &  10366    &    37.97 \\
0300520301 & 2005 Sep 22 &  10034    &    36.33 \\
0311590101 & 2005 Nov 12 &  10917    &    39.57 \\
0400140301 & 2006 May 22 &   5952    &    21.62 \\
0400140401 & 2006 Nov 5  &   5592    &    21.82 \\
0502710201 & 2007 May 5  &   4451    &    16.77 \\
0502710301 & 2007 Nov 17 &   6138    &    24.81 \\
0554510101 & 2009 Mar 21 &   3682    &    15.45 \\
0601170301 & 2009 Sep 22 &   3570    &    15.61 \\
0650920101 & 2011 Apr 11 &   4443    &    19.81 \\
0670700201 & 2011 May 02 &   3741    &    16.99 \\
\hline
total     &              &  140584   &   533.38 \\
\hline
\end{tabular}
\end{table}
The X-ray spectra of \rxj{} are best modelled with a blackbody and a broad absorption feature at $\mathrm{300~eV}$ (probably caused by proton cyclotron resonance) that is represented by a Gaussian absorption line (see \citealt{2009A&A...498..811H,2010A&A...521A..11H,2007Ap&SS.308..181H,2006A&A...451L..17H,2004A&A...415L..31D}), both absorbed due to the interstellar medium. We use the {\it Xspec} model {\it phabs*(bbodyrad+gaussian)}. However, the {\it RGS1} spectrum begins at energies of 0.35~keV, i.e. the centre of the cyclotron line is outside the spectrum (at 0.3~keV) and the information of the continuum yielding the blackbody normalisation, $\mathrm{N_{H}}$ and the properties of the cyclotron line is limited to lower energies. Thus, these parameters are degenerated (a higher $\mathrm{N_{H}}$ value can be compensated by a broader and deeper cyclotron line etc.). For such investigations the {\it EPIC-pn} spectra are much more suitable, but pn does not provide the spectral resolution that is required here.\\
Therefore, we fitted the co-added {\it RGS1} spectrum keeping all parameters free for fitting but fixed the line energy of the cyclotron line at 0.3~keV (also for the error calculations).\\      
We fitted the {\it RGS1} spectrum with this model, using the photons from $\mathrm{0.35-1.0~keV}$. The co-added spectrum fits with $\chi^{2}/d.o.f.=1.50$. For the narrow absorption feature we included a further Gaussian line at $\mathrm{0.57~keV}$. The fit including the narrow absorption feature yields $\chi^{2}/d.o.f.=1.42$. Note, that no instrumental features are present around $\mathrm{0.57~keV}$, i.e. supporting an astrophysical origin. The fit results are listed in \autoref{fit_tab} and the spectrum is shown in \autoref{RGS0720}.\\
We confirm the presence of a narrow absorption feature at $\mathrm{568.6_{-1.9}^{+1.8}~eV}$ with an equivalent width (EW) of $\mathrm{EW=-1.89_{-0.62}^{+0.56}~eV}$ (that corresponds to 5.6$\sigma$ significance).\\

\subsection{\rxd{} (\rbs{})}

\begin{figure}
\centering
\includegraphics*[viewport=105 270 490 560, width=0.425\textwidth]{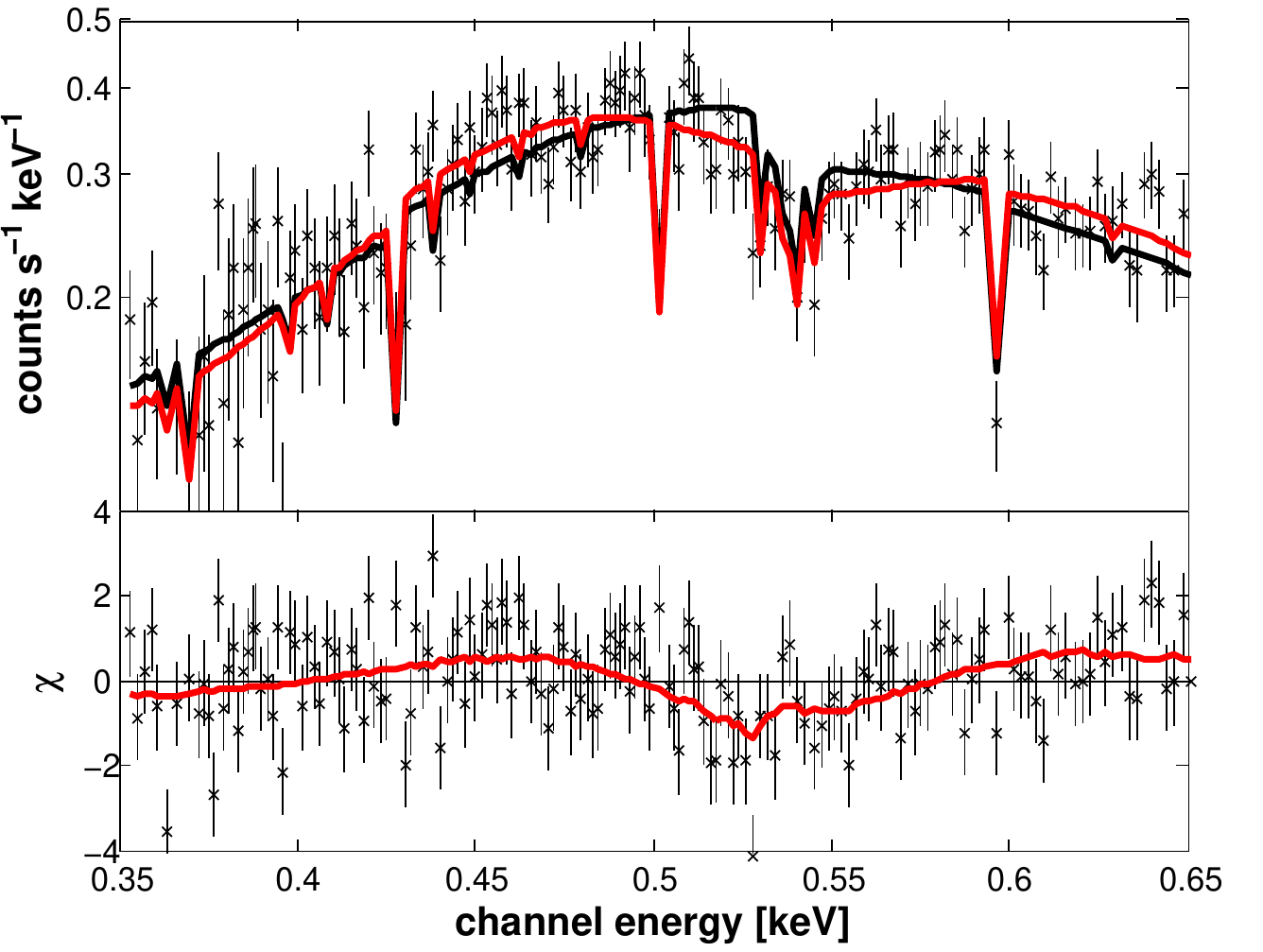}
\caption[]{As in \autoref{RGS0720}, but for \rxd{} with a total exposure of $\mathrm{170~ks}$. The new absorption feature is seen at $\mathrm{535.3_{-13.4}^{+7.4}~eV}$ with $\mathrm{EW=-20_{-13}^{+10}~eV}$ (errors denote $90\%$ confidence level). There is no absorption feature at 0.57~keV}
\label{RGSRBS1223}%
\end{figure}

The \xmm{} {\it EPIC-pn} spectra of the M7 member \rxd{} exhibit two broad Gaussian absorption features \citep{2007Ap&SS.308..181H} with line energies of 0.30~keV and 0.23~keV (or one line at 0.30~keV or 0.29~keV, see \citealt{2003A&A...403L..19H}), again below the {\it RGS} spectrum. Thus, these two lines appear as one broader line with a line energy at 0.205~keV, i.e. we use the {\it Xspec} model {\it phabs*(bbodyrad+gaussian)} and fixed the line energy for fitting, but released it for the error calculation. The \xmm{} {\it RGS} observations of \rxd{} sum up to 190~ks, where 170~ks are left after passing the GTI filters, see \autoref{RGS_1223_tab}.\\
\begin{table}
\centering
\caption[]{As in \autoref{RGS_0720_tab}, but for \rxd{} (\rbs{}).}
\label{RGS_1223_tab}

\begin{tabular}{ccrr}

\hline
Obs. ID & date & counts & eff. exposure \\
	&      & 0.35-1.0~keV & [ks]     \\
\hline

0090010101 & 2001 Dec 31 &   946  &   8.83 \\
0157360101 & 2003 Jan 1  &  3055  &  28.71 \\
0163560101 & 2003 Dec 30 &  3019  &  29.80 \\
0305900201 & 2005 Jun 25 &  1633  &  16.73 \\
0305900301 & 2005 Jun 27 &  1487  &  14.74 \\
0305900401 & 2005 Jul 15 &  1230  &  12.43 \\
0305900601 & 2006 Jan 10 &  1576  &  16.72 \\
0402850301 & 2006 Jun 8 &   545  &   5.53 \\
0402850401 & 2006 Jun 16 &   842  &   8.34 \\
0402850501 & 2006 Jun 27 &   913  &   9.48 \\
0402850901 & 2006 Jul 5 &   662  &   7.11 \\
0402850701 & 2006 Dec 27 &   993  &  10.34 \\

\hline
total     &   & 16901  &  168.76  \\
\hline
\end{tabular}
\end{table}
The co-added spectrum fits with $\chi^{2}/d.o.f.=1.32$. Although having less counts than the co-added spectrum of \rxj{}, it can be clearly stated that no significant narrow absorption feature at 0.57~keV is detected. However, the model overpredicts the count numbers in the energy range from 0.48~keV to 0.60~keV, that can be interpreted as a new absorption feature $-$ broader than the absorption feature in \rxj{}, but too narrow to be detected in {\it EPIC-pn}.\\ 
Including a further (broader) Gaussian absorption line at 0.53~keV decreases $\chi^{2}/d.o.f$ to 1.13. The co-added {\it RGS1} spectrum is shown in \autoref{RGSRBS1223}. \citet{2007Ap&SS.308..619S} reported one absorption feature at 0.20$-$0.39~keV and a further broad feature either at 0.46~keV or at 0.73~keV in the \xmm{} {\it EPIC-pn} spectra of \rxd{}. The second feature does not correspond to the line at 0.53~keV found here and there is no evidence for the feature found in \citet{2007Ap&SS.308..619S} in our co-added {\it RGS1} spectrum. The differences between {\it RGS} and {\it pn} were reduced in the last years (e.g. compared to \citealt{2003A&A...403L..19H} and \citealt{2007Ap&SS.308..619S}) due to new calibrations. Thus an up-dated combined {\it RGS} and {\it pn} analysis is required with the new calibrations.\\
Maybe the X-ray spectrum of \rxd{} is much more complex and the absorption features change with time (since the work of \citealt{2007Ap&SS.308..619S}, we have five more observations available). 
We stress that the {\it RGS} detector exhibits strong features in the energy range from 0.48~keV to 0.60~keV, but these features are narrow compared to the feature we found at 0.53~keV (that is not seen in the brighter spectrum of e.g. \rxj{}). Thus, the new absorption feature is not of an instrumental origin.\\
Since the line centre of the new absorption feature lies at 0.53~keV (the absorption energy of OI at rest, see also~\citealt{2003A&A...404..959D}) we used the {\it Xspec} absorption model {\it vphabs} to vary the abundance of oxygen during the fit procedure. The fit did not find a solution to compensate the absorption feature at 0.53~keV with a higher abundance of neutral oxygen, hence, this feature may originates from the NS itself.\\
The fit results are listed in \autoref{fit_tab}.

\subsection{\rxe{} }

\begin{figure}
\centering
\includegraphics*[viewport=105 270 490 560, width=0.425\textwidth]{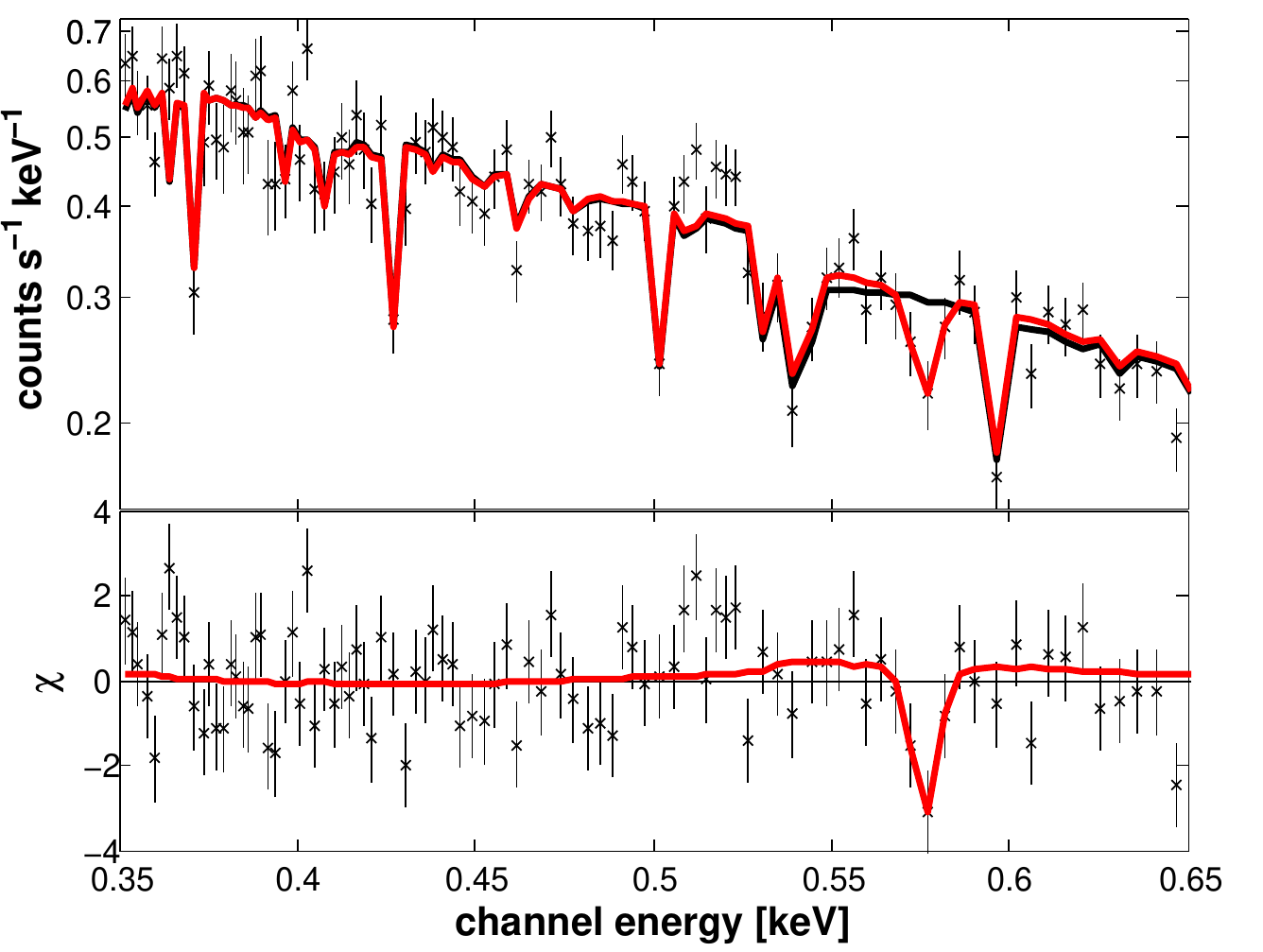}
\caption[]{As in \autoref{RGS0720}, but for \rxe{} with a total exposure of $\mathrm{110~ks}$. The narrow absorption feature is seen at $\mathrm{576.1_{-2.7}^{+2.5}~eV}$ with $\mathrm{EW=-3.2_{-1.8}^{+1.5}~eV}$ (errors denote $90\%$ confidence level), see also \citet{2004ApJ...608..432V} for comparism.}
\label{RGS1605}%
\end{figure}

The first report of a narrow absorption feature in the co-added {\it RGS} spectra of an isolated NS was published by \citet{2004ApJ...608..432V}, who found an absorption line at 0.57~keV in the spectra of \rxe{}. Since then, nine further \xmm{} observations of \rxe{} were performed, i.e. 14 in total. However, almost all observations after February 2006 are strongly contaminated by high background and did not pass the GTI filtering. Finally, our sample of observations is not much different from that used by \citet{2004ApJ...608..432V}, see \autoref{RGS_1605_tab}. We only could add observation 0302140501 (6~ks left after GTI filtering), but omitted observation 0302140601 (used in \citealt{2004ApJ...608..432V}). The total exposure time sums up to 110~ks.\\ 
\begin{table}
\centering
\caption[]{As in \autoref{RGS_0720_tab}, but for \rxe{}.}
\label{RGS_1605_tab}

\begin{tabular}{ccrr}

\hline
Obs. ID & date & counts & eff. exposure \\
	&      & 0.35-1.0~keV & [ks]     \\
\hline

0073140301 & 2002 Jan 9  & 2717     & 20.08 \\
0073140201 & 2002 Jan 15 & 3715     & 27.83 \\
0073140501 & 2002 Jan 19 & 3009     & 22.21 \\
0157360401 & 2003 Jan 17 & 4470     & 30.98 \\
0302140501 & 2006 Feb 12 &  748     & 5.91 \\

\hline
total     &   & 14660   &  107.01  \\
\hline
\end{tabular}
\end{table}
\citet{2004ApJ...608..432V} found a broad absorption feature at 0.493~keV, again associated with a possible cyclotron line, i.e. we use the model {\it phabs*(bbodyrad+gaussian)} here too. This line energy lies well within the {\it RGS} energy range, but the best fit leads to a line energy at 0.403~keV that has to be fixed for the error calculation. Therefore, the $\mathrm{N_{H}}$ value and the blackbody temperature lies below the values listed in \citet{2004ApJ...608..432V}, but is consistent within the errors, see \autoref{fit_tab}. The co-added spectrum fits with $\chi^{2}/d.o.f.=1.22$.\\
Adding a further Gaussian absorption feature, the fit yields $\chi^{2}/d.o.f.=1.11$. We confirm the presence of an absorption line at $\mathrm{576.1_{-2.7}^{+2.5}~eV}$ with $\mathrm{EW=-3.2_{-1.8}^{+1.5}~eV}$, with the 3.5~$\sigma$ significance, as reported in \citet{2004ApJ...608..432V}. The co-added {\it RGS} spectrum of \rxe{} is shown in \autoref{RGS1605}.\\   

\subsection{\rxf{} }
\label{1856}

\begin{figure}
\centering
\includegraphics*[viewport=105 270 490 560, width=0.425\textwidth]{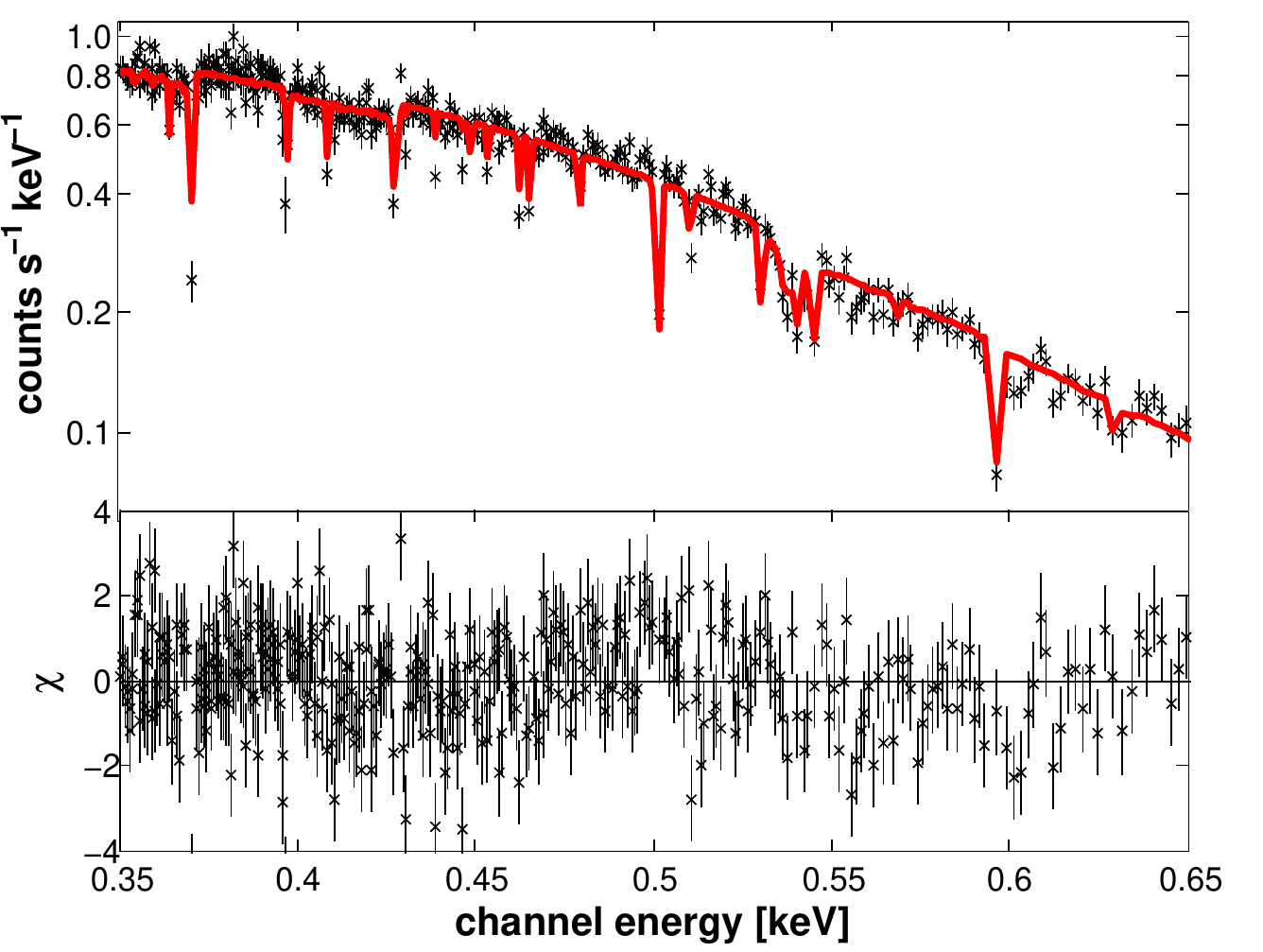}
\caption[]{As in \autoref{RGS0720}, but for \rxf{} with a total exposure of $\mathrm{880~ks}$. The spectrum of \rxf{} exhibits no spectral lines.}
\label{RGS1856}%
\end{figure}

This NS is the brightest member of the M7 and is known to exhibit no features in \xmm{} {\it EPIC-pn} and {\it Chandra HRC-S/LETG} spectra \citep{2001A&A...379L..35B,2003A&A...399.1109B} and is constant over time. Thus, \rxf{} is used as calibration source for X-ray telescopes. Therefore, the \xmm{} observations sum up to 1000~ks, where 880~ks remain after the GTI filtering, see \autoref{RGS_1856_tab}.
\begin{table}
\centering
\caption[]{As in \autoref{RGS_0720_tab}, but for \rxf{}.}
\label{RGS_1856_tab}

\begin{tabular}{ccrr}

\hline
Obs. ID & date & counts & eff. exposure \\
	&      & 0.35-1.0~keV & [ks]     \\
\hline

0106260101 & 2002 Apr 8	 & 8417  &    57.30 \\
0201590101 & 2004 Apr 17 & 5158	 &    36.61 \\
0165971701 & 2004 Sep 24 & 4742	 &    33.73 \\
0165971601 & 2004 Sep 24 & 4654	 &    33.29 \\
0165971901 & 2005 Mar 23 & 2531	 &    17.91 \\
0213080101 & 2005 Apr 15 & 1579  &    11.39 \\
0165972001 & 2005 Sep 24 & 4624	 &    33.53 \\
0165972101 & 2006 Mar 26 & 8953	 &    68.71 \\
0412600101 & 2006 Oct 24 & 8715  &    72.75 \\
0412600201 & 2007 Mar 14 & 6049  &    50.24 \\
0415180101 & 2007 Mar 25 & 3042  &    22.72 \\
0412600301 (S004) & 2007 Oct 4 & 2926     &	 22.25 \\
0412600301 (U002) & 2007 Oct 4 & 2459     &	 19.69 \\
0412600401 & 2008 Mar 13 & 6052  &    48.34 \\
0412600601 & 2008 Oct 4	 & 7913  &    64.97 \\
0412600701 & 2009 Mar 19 & 8300  &    68.65 \\
0412600801 (S004) & 2009 Oct 7 &  872	  &	 6.90 \\
0412600801 (U002) & 2009 Oct 7 & 8576	  &	70.88 \\
0412600901 & 2010 Mar 22 & 8652  &    72.22 \\
0412601101 & 2010 Sep 28 & 7950  &    68.30 \\

\hline
total     &     & 112170   &  880.38   \\
\hline
\end{tabular}
\end{table}
Although \citet{2001A&A...379L..35B,2003A&A...399.1109B} found no absorption feature we include this source as well, since the co-added {\it RGS} spectra of \rxf{} were not investigated before.\\
Likewise \citet{2001A&A...379L..35B,2003A&A...399.1109B}, we apply the {\it Xspec} model {\it phabs*(bbodyrad)}, where the fit yields $\chi^{2}/d.o.f.=1.16$ (see \autoref{fit_tab}). The spectrum of \rxf{} is completely featureless in the {\it RGS1} energy range (\autoref{RGS1856}).    

\subsection{Geminga (PSR\,J0633+1746) }

\begin{figure}
\centering
\includegraphics*[viewport=105 270 490 560, width=0.425\textwidth]{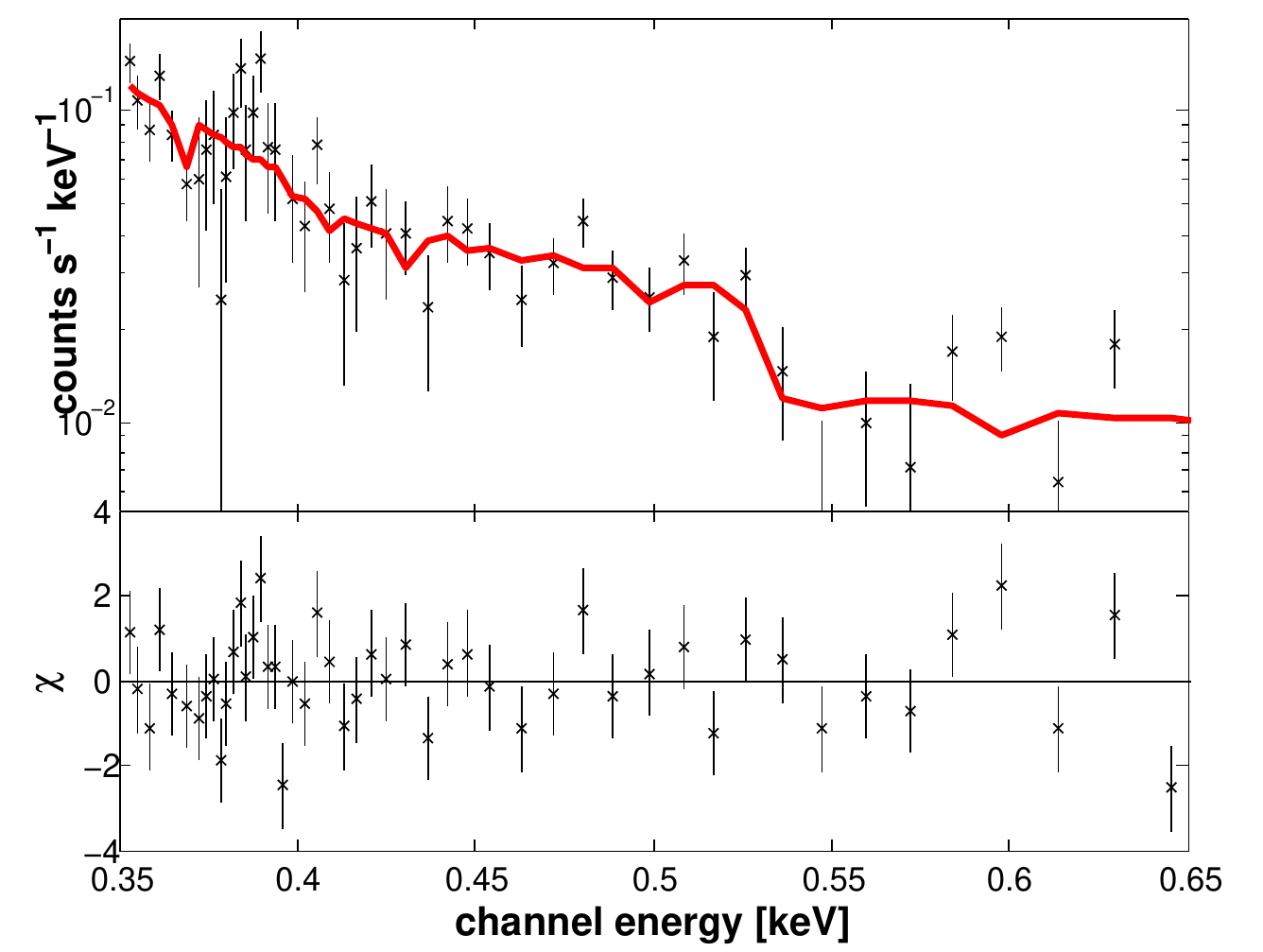}
\caption[]{As in \autoref{RGS0720}, but for Geminga (with a total exposure of $\mathrm{240~ks}$) fitted with one single blackbody and an additive power-law component. The spectrum of Geminga exhibits no lines.}
\label{Geminga}%
\end{figure}

The first spectrum of Geminga with a sufficiently large count number was published by \citet{1993ApJ...415..286H}, using {\it ROSAT} data. They found the Geminga spectrum is best modelled with two blackbodies having $\mathrm{kT_{eff}=43~eV}$ and $\mathrm{kT_{eff}=260~eV}$, respectively. Using new  {\it ROSAT} data, \citet{1997ApJ...477..905H} identified a power-law component above 2~keV with $\mathrm{\Gamma=0.5-1.5}$ (the most recent value is $\mathrm{\Gamma=1.72\pm0.10}$, see \citealt{2002ApJ...578..935J,2004Sci...305..376C}). Therefore, we use {\it phabs*(bbodyrad+bbodyrad+pow)} for the data fitting.\\
Although Geminga was frequently observed with \xmm{} (240~ks after GTI filtering), the total number of counts is only 3,100, i.e. less than half the photons \citet{1993ApJ...415..286H} could use in their first Geminga spectrum (but with lower spectral resolution), see \autoref{RGS_Geminga_tab}.\\
\begin{table}
\centering
\caption[]{As in \autoref{RGS_0720_tab}, but for Geminga.}
\label{RGS_Geminga_tab}

\begin{tabular}{ccrr}

\hline
Obs. ID & date & counts & eff. exposure \\
	&      & 0.35-2.0~keV & [ks]     \\
\hline

 0111170101 & 2002 Apr 4  & 993     & 76.38 \\
 0201350101 & 2004 Mar 13 & 259     & 15.85 \\
 0301230101 & 2005 Sep 16 &  49     &  4.91 \\
 0311591001 & 2006 Mar 17 & 521     & 34.80 \\
 0400260201 & 2007 Oct 2  & 228     & 20.17 \\
 0400260301 & 2007 Mar 11 & 277	    & 21.71 \\
 0501270201 & 2007 Sep 18 & 211     & 22.73 \\
 0501270301 & 2008 Mar 8  & 127     & 11.82 \\
 0550410201 & 2008 Oct 3  & 240     & 21.11 \\
 0550410301 & 2009 Mar 10 & 181     & 11.36 \\

\hline
total      &   & 3086    & 240.84 \\
\hline
\end{tabular}
\end{table}
Since the co-added {\it RGS} spectrum of Geminga has the lowest count number in our sample and the hot blackbody dominates the spectrum above 1~keV, but is strongly superimposed by the power-law (see e.g. Figure~1 in \citealt{2005ApJ...623.1051D}). We find the best fit with a negligible normalisation of the hot blackbody, whereas the power-law is clearly visible. Thus, we applied the model {\it phabs*(bbodyrad+pow)}. The fit yields $\chi^{2}/d.o.f.=1.21$ and the results are shown in \autoref{fit_tab}.\\
We find no evidence for an absorption feature like in the case of \rxj{}, \rxd{} or \rxe{} in the co-added {\it RGS} spectrum of Geminga. However, weak features cannot be detected due to the low count number (\autoref{Geminga}).

\subsection{PSR\,B0656+14}

\begin{figure}
\centering
\includegraphics*[viewport=105 270 490 560, width=0.425\textwidth]{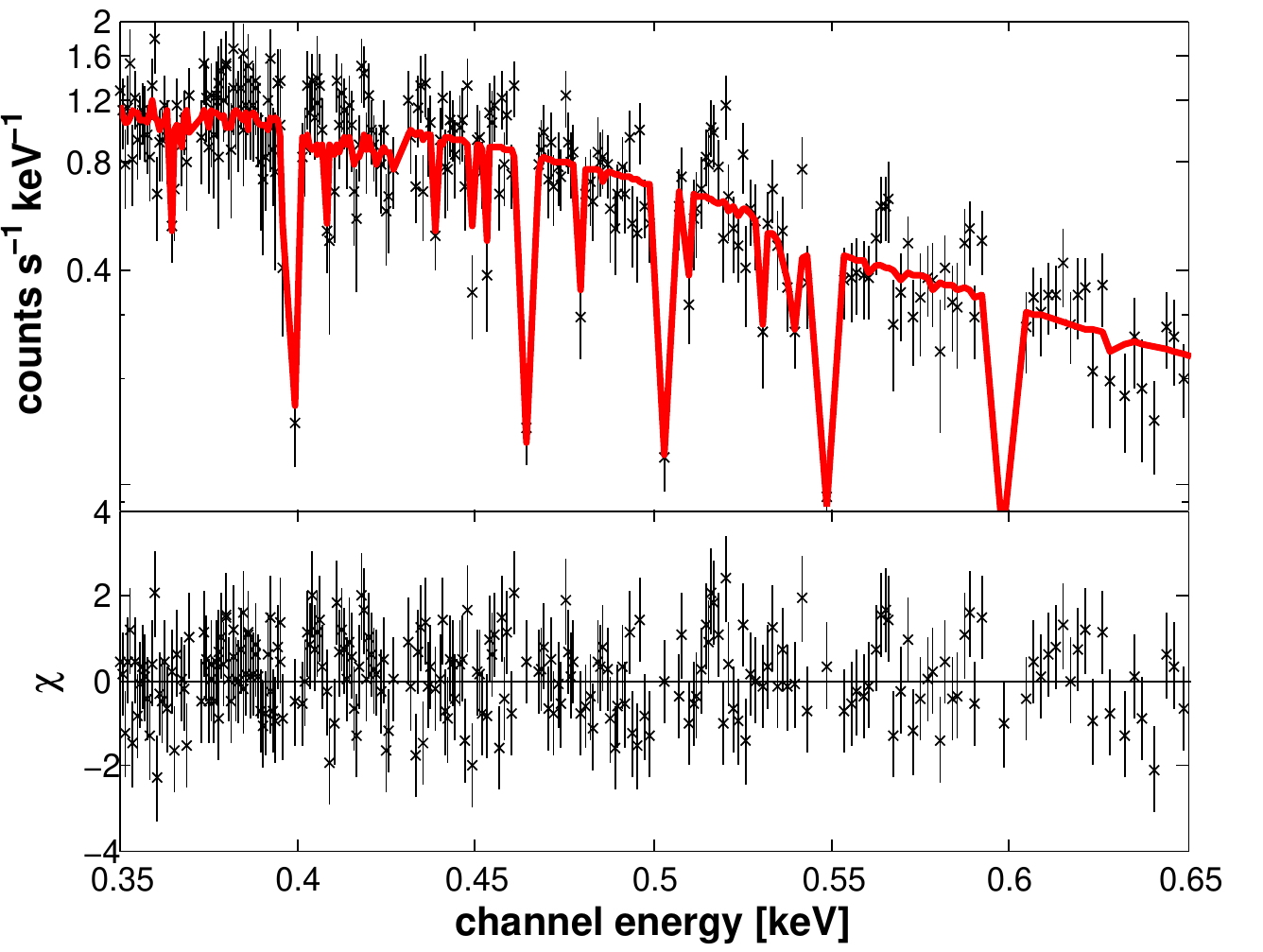}
\caption[]{As in \autoref{RGS0720}, but for PSR\,B0656+14 with a total exposure of $\mathrm{38~ks}$. The spectrum of PSR\,B0656+14 exhibits no lines.}
\label{PSRB0656}%
\end{figure}

The first investigation of the emission of PSR\,B0656+14 was performed with {\it ROSAT} data by \citet{1996A&A...313..565P}, who modelled the spectrum with a blackbody and a second component that could be either a further blackbody or a power-law. Later, \citet{1996ApJ...465L..35G} applied the three component model (two blackbodies and a power-law with $\mathrm{\Gamma=1.5\pm1.1}$) that yields the best fit. This model was later confirmed by \citet{2002nsps.conf..273P} using {\it Chandra} data. \citet{2002ApJ...574..377M} found no evidence for spectral features in the {\it Chandra} spectra of PSR\,B0656+14, but used only two blackbodies (and no power-law) for their fit model. Today, only one {\it RGS} observation of PSR\,B0656+14 is available (\autoref{RGS_0656_tab}). But this observation yields more counts than e.g. all Geminga observations in total.\\
\begin{table}
\centering
\caption[]{As in \autoref{RGS_0720_tab}, but for PSR\,B0656+14.}
\label{RGS_0656_tab}

\begin{tabular}{ccrr}

\hline
Obs. ID & date & counts & eff. exposure \\
	&      & 0.35-2.0~keV & [ks]     \\
\hline
0112200101 & 2001 Oct 23 &  7483  &  38.12 \\
\hline
\end{tabular}
\end{table}
We first fitted the {\it RGS1} spectrum of PSR\,B0656+14 with {\it phabs*(bbodyrad+bbodyrad+pow)}. Both blackbodies dominate the spectrum in the {\it RGS} energy range, but the power-law becomes significant above 2~keV (see Figure~2 in \citealt{2005ApJ...623.1051D}), where the {\it RGS1} spectrum has much less counts. Thus, the power-law is negligible and we further on use the  model {\it phabs*(bbodyrad+bbodyrad)} (likewise in \citealt{2002ApJ...574..377M}). The fit yields $\chi^{2}/d.o.f.=0.88$ and the spectral properties are listed in \autoref{fit_tab}.\\
We find no evidence for a narrow absorption feature in the {\it RGS1} spectrum of PSR\,B0656+14 (\autoref{PSRB0656}) and thereby confirm the results in \citet{2002ApJ...574..377M}. The count number of the {\it RGS1} spectrum would have been sufficiently high to detect an absorption feature as found e.g. in the spectrum of \rxj{}, if present.      

\subsection{PSR\,B1055-52}

\begin{figure}
\centering
\includegraphics*[viewport=100 270 490 560, width=0.425\textwidth]{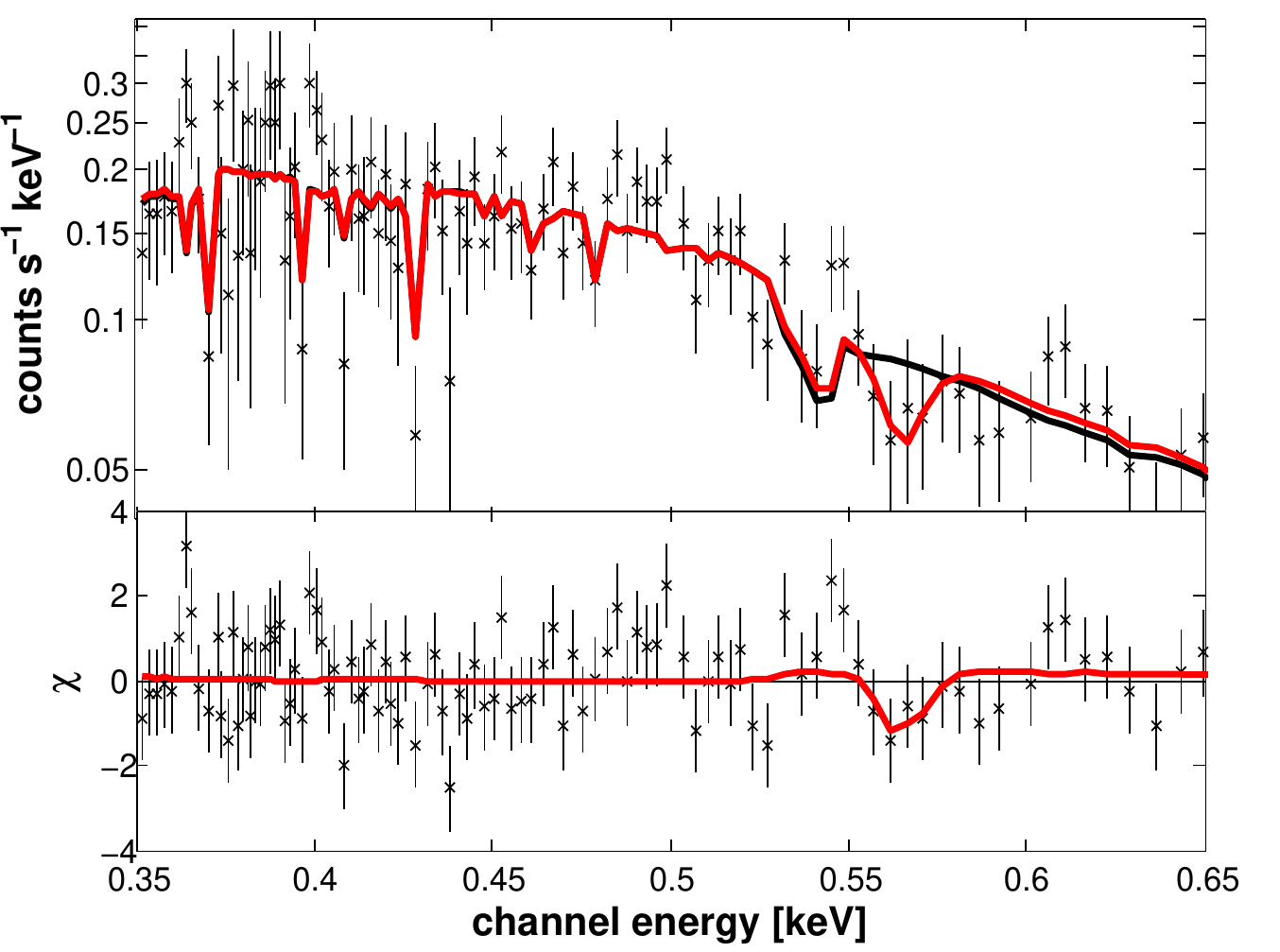}
\caption[]{As in \autoref{RGS0720}, but for PSR\,B1055-52 with a total exposure of $\mathrm{90~ks}$. The spectrum of PSR\,B1055-52 exhibits a new absorption feature seen at $\mathrm{565.0_{-7.3}^{+26.0}~eV}$ with $\mathrm{EW=-5.7_{-9.2}^{+4.6}~eV}$ (errors denote $90\%$ confidence level)}
\label{PSRB1055}%
\end{figure}

The first spectral investigation of PSR\,B1055-52 was performed by \citet{1993ApJ...413L..31O} using {\it ROSAT} data. Like in the case of PSR\,B0656+14, the best fit model required a blackbody and a further component that could be a second blackbody or a power-law with $\mathrm{\Gamma\approx4}$. Combining {\it ROSAT} and {\it Chandra} data, \citet{2002nsps.conf..273P} showed that an adequate fit model requires three components: two blackbodies and a power-law with $\mathrm{\Gamma\approx1.7}$. Therefore we first use {\it phabs*(bbodyrad+bbodyrad+pow)}.\\
Altogether, the {\it RGS1} data of PSR\,B1055-52 sum up to 95~ks (90~ks after GTI filtering) with $\approx$4,000 photons (\autoref{RGS_1055_tab}).
\begin{table}
\centering
\caption[]{As in \autoref{RGS_0720_tab}, but for PSR\,B1055-52.}
\label{RGS_1055_tab}

\begin{tabular}{ccrr}

\hline
Obs. ID & date & counts & eff. exposure \\
	&      & 0.35-2.0~keV & [ks]     \\
\hline

0113050101 & 2000 Dec 14 &  1109   &  22.75  \\
0113050201 & 2000 Dec 15 &  2323   &  55.66  \\
0113050801 & 2000 Dec 15 &   164   &   3.77  \\
0113050901 & 2000 Dec 15 &   356   &   7.84  \\

\hline
total     &     &  3952  &  90.02   \\
\hline
\end{tabular}
\end{table}
For the same reasons as discussed in the case of Geminga, we found the best fit for a negligible normalisation of the second, hotter, blackbody. Therefore, we applied the model {\it phabs*(bbodyrad+pow)}, that fits with $\chi^{2}/d.o.f.=1.16$ (\autoref{fit_tab}).\\
In contrast to the other 3M, PSR\,B1055-52 seems to exhibit a weak absorption feature at 0.57~keV, similar to those found in the co-added {\it RGS1} spectra of \rxj{} and \rxe{}. To check this, we add a Gaussian absorption line to our fit model (\autoref{PSRB1055}). With the new model, the $\chi^{2}/d.o.f.$ value does not change significantly to $\chi^{2}/d.o.f.=1.15$, but the fit easily finds its minimum with the new line at $\mathrm{565.0_{-7.3}^{+26.0}~eV}$ with $\mathrm{EW=-5.7_{-9.2}^{+4.6}~eV}$. This feature is not consistent to the continuum within 2$\sigma$, i.e. less significant than the narrow absorption feature in \rxe{}, but we stress that the total count number in the case of the co-added {\it RGS1} spectrum of PSR\,B1055-52 is more than three times less than in the case of \rxe{}, i.e. further observations are required to verify this absorption feature.       %
\begin{sidewaystable*}
\centering
\caption[]{Fit results from the co-added \xmm{} {\it RGS} spectra of four M7 and the 3M. The sizes of the emitting areas ({\it norm}) are calculated from the normalisations of the X-ray blackbodies and correspond to a distance of 300~pc, except in the case of \rxf{} (130~pc, \citealt{2010arXiv1008.1709W}) and PSR\,B1055-52 (750~pc, \citealt{2003MNRAS.342.1299K}). The line energies of the broad absorption features are always fitted as free parameters (except in the case of \rxj{}, where the line energy was always fixed), but fixed for the error calculation. The equivalent widths (EW) of these broad absorption features have no errors since the lines reach partly out of the energy range of the {\it RGS} spectrum. All errors denote 90\% confidence level.}
\label{fit_tab}

\begin{tabular}{ll|ccccccc}

\hline
model  &    unit                              &    \rxj{}                       & \rxd{}                          & \rxe{}                          &   \rxf{}                         & Geminga                          & PSR\,B0656+14                   & PSR\,B1055-52\\
component  & 	                              &                                 &                                 &                                 &                                  &                                  &                                 &            \\
\hline
$\mathrm{N_{H}}$ & $\mathrm{10^{20}/cm^{2}}$  & $\mathrm{1.60^{+0.52}_{-0.65}}$ & $\mathrm{1.6_{-1.5}^{+1.6}}$    & $\mathrm{0.72_{-0.72}^{+1.72}}$ &   $\leq0.32$                     & $\mathrm{3.9_{-1.9}^{+11.1}}$    & $\mathrm{0.12_{-0.12}^{+1.64}}$ & $\mathrm{2.4_{-2.4}^{+3.4}}$  \\
\hline
$\mathrm{bbodyrad_{1}}$&                      &                                 &                                 &                                 &                                  &                                  &                                   &            \\
$\mathrm{kT_{1}}$& eV                         & $\mathrm{85.0_{-1.6}^{+1.7}}$   & $\mathrm{78.3_{-3.3}^{+3.6}}$   & $\mathrm{85.3_{-4.2}^{+3.1}}$   & $\mathrm{62.67_{-0.60}^{+0.25}}$ & $\mathrm{33.2_{-12.1}^{+9.2}}$   &  $\mathrm{67.9_{-6.7}^{+3.6}}$    & $\mathrm{72.0_{-7.8}^{+9.5}}$    \\
$\mathrm{norm_{1}}$& km                       & $\mathrm{6.47_{-0.61}^{+0.64}}$ & $\mathrm{6.7_{-1.4}^{+1.8}}$    & $\mathrm{4.81_{-0.72}^{+1.44}}$ &$\mathrm{5.015_{-0.035}^{+0.260}}$& $\leq68$                        & $\mathrm{9.53_{-0.72}^{+4.66}}$   &$\mathrm{8.8_{-4.0}^{+4.3}}$   \\
\hline
$\mathrm{bbodyrad_{2}}$&                      &     $-$                         &     $-$                         &     $-$                         &     $-$                          &     $-$                          &                                   &     $-$            \\
$\mathrm{kT_{2}}$& eV                         &     $-$                         &     $-$                         &     $-$                         &     $-$                          &     $-$                          & $\mathrm{170_{-44}^{+36}}$       &     $-$            \\
$\mathrm{norm_{2}}$& km                       &     $-$                         &     $-$                         &     $-$                         &     $-$                          &     $-$                          & $\mathrm{0.373_{-0.088}^{+0.608}}$&     $-$            \\
\hline
$\mathrm{Gaussian_{1}}$& 	              &                                 &                                 &                                 &     $-$                          &     $-$                           &     $-$                           &     $-$            \\
$\mathrm{lineE_{1}}$ & eV	              &   300 (fixed)                   &  205.83 (fixed)                 &  403 (fixed)                    &     $-$                          &     $-$                           &     $-$                           &     $-$            \\
$\mathrm{\sigma_{1}}$ width& eV                    & $\mathrm{100.7_{-12.4}^{+8.5}}$ & $\mathrm{135.7_{-3.2}^{+2.5}}$  & $\mathrm{100.1_{-5.7}^{+7.0}}$  &     $-$                          &     $-$                           &     $-$                           &     $-$            \\
$\mathrm{EW_{1}}$& eV                         &   -35.2                         &  -163                           &  -65.2                          &     $-$                          &     $-$                           &     $-$                           &     $-$            \\
\hline                        
power law    &                                &     $-$                         &     $-$                         &     $-$                         &     $-$                          &                                  &   $-$                           &            \\
$\mathrm{\Gamma}$ &                           &     $-$                         &     $-$                         &     $-$                         &     $-$                          & $\mathrm{2.67_{-0.91}^{+1.75}}$  &   $-$                           &$\mathrm{3.8_{-2.4}^{+1.8}}$   \\
 norm & ph $\mathrm{/keV/cm^{2}/s}$   &     $-$                         &     $-$                         &     $-$                         &     $-$                          & $\mathrm{9.1_{-2.3}^{+4.5}\times10^{-5}}$ &   $-$                           & $\mathrm{7.6_{-4.3}^{+4.0}\times10^{-5}}$            \\
 &  $\mathrm{@ 1~keV}$ & & & & & & & \\
\hline
$\mathrm{\chi^{2}_{red}/d.o.f.}$ &            & 1.50/456                        &  1.32/208                       &  1.22/110                       &  1.16/759                        &  1.21/66  &  0.88/305                        & 1.16/121          \\ 
\hline
\hline
$\mathrm{Gaussian_{2}}$& 	              &                                 &                                 &                                 &     $-$                          &     $-$                           &   $-$                           &            \\
$\mathrm{lineE_{2}}$ & eV	              & $\mathrm{568.6_{-1.9}^{+1.8}}$  & $\mathrm{535.3_{-13.4}^{+7.4}}$ & $\mathrm{576.1_{-2.7}^{+2.5}}$  &     $-$                          &     $-$                            &   $-$                           & $\mathrm{565.0_{-7.3}^{+26.0}}$  \\
$\mathrm{\sigma_{2}}$ width& eV                    & $\mathrm{4.8_{-1.9}^{+1.8}}$    & $\mathrm{35_{-18}^{+21}}$       & $\mathrm{3.3_{-3.3}^{+4.1}}$    &     $-$                          &     $-$                            &   $-$                           & $\mathrm{5.6_{-5.6}^{+20.1}}$    \\
$\mathrm{EW_{2}}$& eV                         & $\mathrm{-1.89_{-0.62}^{+0.56}}$& $\mathrm{-20_{-13}^{+10}}$      & $\mathrm{-3.2_{-1.8}^{+1.5}}$   &     $-$                          &     $-$                            &   $-$                           & $\mathrm{-5.7_{-9.2}^{+4.6}}$   \\
significance & $\sigma$ & 5.6 & 3.2 & 3.5 & $-$ & $-$ & $-$ & 2.1 \\
\hline
$\mathrm{\chi^{2}_{red}/d.o.f.}$ &            & 1.42/453                        & 1.13/205                        & 1.11/107                        &     $-$                          &     $-$                            &   $-$                           & 1.15/118          \\ 
\hline
\end{tabular}
\end{sidewaystable*}
%
\subsection{ \xmm{} Results}

We co-added all available {\it RGS} data of the four brightest M7 (\rxj{}, \rxd{}/\rbs{}, \rxe{} and \rxf{}) and the 3M; Geminga, PSR\,B0656+14 and PSR\,B1055-52. We confirm the narrow absorption features at 0.57~keV in the spectra of \rxj{} and \rxe{} as reported in earlier work by \citet{2009A&A...497L...9H} and \citet{2004ApJ...608..432V}, respectively. We found a new absorption feature at 0.53~keV in the co-added {\it RGS} spectrum of \rxd{}, that is broader than those found in the case of \rxj{} and \rxe{}.\\
The co-added spectra of \rxf{} and PSR\,B0656+14 are featureless, while the count number in the co-added spectra of Geminga is not sufficiently high to exclude such weak features. There might be a new absorption feature (comparable to those in the spectra of \rxj{} and \rxe{}) present in the co-added spectrum of PSR\,B1055-52.\\
We fitted the absorption features using an additive Gaussian line and checked our results with the multiplicative \textit{gabs} model. For both models we obtain exact the same values for the equivalent widths.\footnote{The documentation of \textit{gabs} in the \textit{Xspec} manual is misleading and might cause some confusion. The $\mathrm{\tau}$ (parameter 3) in \textit{gabs} is not the optical depth (although it is even called so in older manuals). The ``correct" optical depth can be calculated by $\mathrm{\tau=\tau_{Xspec}/P_{2}/\sqrt{2 \pi}}$, where $\mathrm{P_{2}}$ is the second parameter in the  \textit{gabs} model. This should be taken into account if equivalent widths have to be calculated. We thank Oleg Kargaltsev, who pointed this out (see also \textit{http://xspector.blogspot.com/2011/07/note-on-gabs-model.html}).}\\
Our results are summarised in~\autoref{fit_tab} and we will discuss them in the next sections.   
%
%
%
%
%
%
\section{The {\it Chandra HRC-S/LETG} spectra of \rxj{} and \rxf{}}

\citet{2009A&A...497L...9H} considered the absorption feature at 0.57~keV could be caused by gravitational redshifted OVIII. If true, OVII should appear at 0.48~keV with the same redshift ($\mathrm{g_{r}=1.17}$, see the more detailed discussion in~\autoref{conclusions}). However, we found no evidence for such a feature in the co-added {\it RGS} spectrum, see~\autoref{RGS0720}. Therefore we also analysed the {\it Chandra HRC-S/LETG} spectra of \rxj{}\footnote{\rxe{} and \rxd{}(\rbs{}) are not as frequently observed as \rxj{} with {\it Chandra HRC-S/LETG}. For example the 90~ks {\it HRC-S/LETG} observation of \rxd{} has four times less counts than the co-added {\it RGS} spectrum. Thus, we do not discuss these spectra in this work.} and \rxf{} for comparison, since \rxf{} does not exhibit any features in the {\it Chandra HRC-S/LETG} spectra \citep{2001A&A...379L..35B,2003A&A...399.1109B}.\\
The {\it Chandra HRC-S/LETG} \citep{1996ChNew...4....9J,1997ChNew...5...16E,1997ChNew...5...20J,1997SPIE.3114...53K} data were analysed with CIAO 4.1. We created own GTI files using {\it dmgti} and filtered those events for data reduction with less than 60$-$180~cts/s (depending on observation and background lightcurves). The photons (source plus background and background) from both first orders were cut out within the standard {\it LETG} spectral extraction windows. The {\it HRC-S/LETG} spectra were added using the CIAO command {\it add~grating~orders} to add the two first orders (all other orders have a negligible count number) and {\it add~grating~spectra} to add the {\it HRC-S/LETG} spectra. To prepare the {\it HRC-S/LETG} data for spectroscopic fitting in {\it Xspec}, the background was generated with the command {\it tg$-$bkg}. Altogether, the co-added {\it HRC-S/LETG} spectra have $\mathrm{429~ks}$ exposure time (see~\autoref{chandraobs}). For details of the reduction of the {\it Chandra} data we also refer to \citet{2010A&A...521A..11H}.\\
\begin{table}
   \centering
     \caption[]{{\it Chandra HRC-S/LETG} observations of \rxj\ in chronological order. We list the exposure times and the net counts after the data passed the GTI filters.}
        \label{chandraobs}

{\footnotesize
\begin{tabular}{lr|crr}

\hline
Obs. Id.	&   	     Start Date   & Counts  & Eff. exposure \\    
		      &      		      	      & 	0.15-1.0~keV             & [ks]      \\
\hline

368   &  2000 Feb  1 &       580		& 2.10 \\
745   &  2000 Feb  2 &       3   		& 0.14  \\
369   &  2000 Feb  4 &       2660		& 6.12 \\
5305  &  2004 Feb 27 &       9691		& 35.68 \\  
5581  &  2005 Jan 23 &       17703  & 62.05 \\    
5582  &  2005 Jun  1 &       19639  & 69.69 \\
6364  &  2005 Aug 27 &       9323 	& 31.68 \\	     	     
6369  &  2005 Oct  8 &       4448		& 16.14 \\
7177  &  2005 Oct  9 &       2067		&  7.18 \\	      
7243  &  2005 Dec 14 &       4992		& 17.05 \\
7244  &  2005 Dec 15 &       4295		& 16.16 \\
7245  &  2005 Dec 16 &       2964		& 11.10 \\
5584  &  2005 Dec 17 &       2972		& 10.87 \\	     
7251  &  2006 Sep 9  &       2713		& 10.59 \\
10861 &  2009 Jan 20 &       2948		& 11.11 \\		   
10700 &  2009 Feb 14 &       5478		& 21.61 \\		
10701 &  2009 Sep 11 &       8228		& 32.36 \\
11820 &  2010 Jun 19 &       8650		& 33.56 \\
13181 &  2010 Nov 18 &       4937		& 19.90 \\
13188 &  2010 Nov 19 &       3709		& 14.04 \\
  
\hline
total & 	     &        118000    	       & 429.60 \\																      
\hline																       
\end{tabular}
}
\end{table}
We fitted the co-added {\it Chandra HRC-S/LETG} spectrum of \rxj{} with the same model as in~\autoref{fit_tab} (without the narrow absorption feature at 0.57~keV), using the photons from $0.15-1.0~\mathrm{keV}$, but ignoring the edges of the chip gaps at $0.2390-0.2432~\mathrm{keV}$, $0.248-0.252~\mathrm{keV}$ and $0.2180-0.2215~\mathrm{keV}$, to avoid systematic fit errors. The fit yields $\chi^{2}/d.o.f.=1.07$ for 258 degrees of freedom applying the standard model with blackbody radiation and the broad absorption feature at $0.3~\mathrm{keV}$. The spectral properties are listed in~\autoref{bbfitHRC} and are consistent to those derived from \xmm{} {\it EPIC-pn} spectra, see \citet{2009A&A...498..811H}\footnote{The {\it EPIC-pn} spectra are fitted for the individual observations in one session. Since \rxj{} is variable, only $\mathrm{N_{H}}$, line energy and line $\sigma$ are kept constant in time.} and~\autoref{fit_tab}. Note, the averaged temperature of \rxj{} derived from the {\it Chandra HRC-S/LETG} spectrum (\autoref{bbfitHRC}) is different to that obtained with the {\it RGS} data (\autoref{fit_tab}), since the source is variable and the observations of the different instruments were performed at different times. In addition, possible systematic errors in the effective area calibrations between both instruments can lead to different temperatures.\\
\begin{figure}
\centering
\includegraphics*[viewport=95 65 595 455, width=0.425\textwidth]{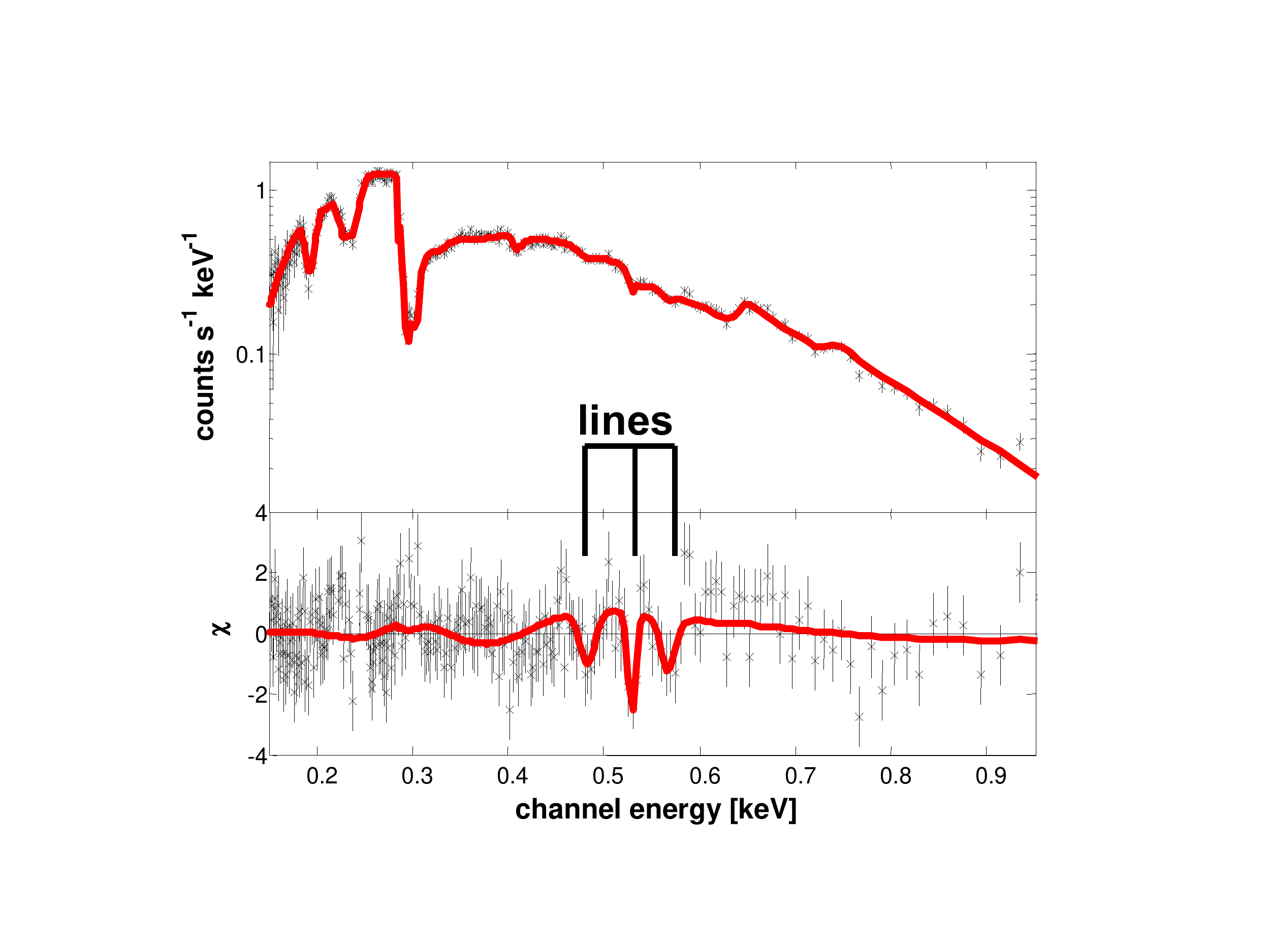}
\caption[]{The co-added {\it Chandra HRC-S/LETG} spectrum of \rxj{} with a total exposure of $\mathrm{429~ks}$. The black line represents the fit model, while the red line shows the fit model plus three narrow absorption feature seen at 0.48~keV, 0.53~keV and 0.57~keV, see also~\autoref{bbfitHRC}.}
\label{RXJ0720_LETG}%
\end{figure}
\begin{table*}
\centering
\caption[]{Fit results from the co-added {\it Chandra HRC-S/LETG} spectrum of \rxj{} (blackbody with Gaussian absorption lines, where EW denotes the equivalent widths) and \rxf{} (blackbody). The sizes of the emitting areas ({\it norm}) are calculated from the normalisations of the X-ray blackbodies and correspond to a distance of 300~pc for \rxj{} \citep{2007ApJ...660.1428K,EisenPhd} and 130~pc for \rxf{} \citep{2010arXiv1008.1709W}. All errors denote $90\%$ confidence level.}
\label{bbfitHRC}
\begin{tabular}{llcc}
\hline
 model               & unit                           & \rxj{}                             & \rxf{} \\
component            &                                &                                    &        \\ 
\hline
$\mathrm{N_{H}}$     &$\mathrm{10^{20}/cm^{2}}$       & $\mathrm{0.886_{-0.047}^{+0.048}}$ &  $\mathrm{0.800_{-0.032}^{+0.033}}$      \\
\hline
bbodyrad             &                                &                                    &        \\ 
kT                   & eV                             & $\mathrm{92.4_{-1.0}^{+0.9}}$      & $\mathrm{63.28\pm0.48}$\\
norm                 & km                             & $\mathrm{4.5_{-1.1}^{+1.3}}$       & $\mathrm{4.95\pm0.12}$         \\
\hline
$\mathrm{Gaussian_{1}}$&                              &                                    &        \\ 
$\mathrm{lineE_{1}}$& eV                              & $\mathrm{293.4_{-6.9}^{+5.0}}$     &   $-$     \\
$\mathrm{\sigma_{1}}$ width& eV                             &  $\mathrm{50.45_{-7.1}^{+9.0}}$    &   $-$     \\
$\mathrm{EW_{1}}$   & eV                              & $\mathrm{-29.6_{-4.6}^{+4.7}}$  &      $-$  \\
\hline
$\mathrm{\chi^{2}_{red}}$/d.o.f. &  & 1.07/258& 1.02/505\\
\hline
\hline
$\mathrm{Gaussian_{2}}$&                              &                                    &        \\ 
$\mathrm{lineE_{2}}$& eV                              & $\mathrm{482.6_{-40.5}^{+5.1}}$     &   $-$     \\
$\mathrm{\sigma_{2}}$ width& eV                             &  $\mathrm{7.8_{-4.9}^{+9.0}}$    &     $-$   \\
$\mathrm{EW_{2}}$    & eV                             & $\mathrm{-1.83_{-2.76}^{+0.98}}$  &    $-$    \\
significance & $\sigma$ & 3.0 &  $-$    \\
\hline
$\mathrm{Gaussian_{3}}$&                              &                                    &        \\ 
$\mathrm{lineE_{3}}$ & eV                             & $\mathrm{529.4_{-2.7}^{+2.2}}$     &    $-$    \\
$\mathrm{\sigma_{3}}$ width& eV                             &  $\mathrm{3.8_{-1.6}^{+2.9}}$    &     $-$   \\
$\mathrm{EW_{3}}$    & eV                             & $\mathrm{-1.79_{-0.84}^{+0.74}}$  &     $-$   \\
significance & $\sigma$ & 3.9 &  $-$    \\
\hline
$\mathrm{Gaussian_{4}}$ &                              &                                    &       \\ 
$\mathrm{lineE_{4}}$ & eV                              & $\mathrm{566.4_{-7.2}^{+6.6}}$     &    $-$    \\
$\mathrm{\sigma_{4}}$ width & eV                             &  $\mathrm{7.4_{-4.0}^{+5.3}}$    &    $-$    \\
$\mathrm{EW_{4}}$    & eV                              & $\mathrm{-2.0_{-1.3}^{+1.2}}$  &    $-$    \\
significance & $\sigma$ & 2.7 &  $-$    \\
\hline
$\mathrm{\chi^{2}_{red}}$/d.o.f. &  & 1.02/249 & \\
\hline
\end{tabular}
\end{table*}%
The narrow absorption feature at $\mathrm{0.57~keV}$ found by \citet{2009A&A...497L...9H} should also be present in the co-added {\it Chandra HRC-S/LETG} spectrum of \rxj{}. Indeed, regarding the fit residuals in~\autoref{RXJ0720_LETG}, such an absorption feature appears at the particular energy range, but less clear than in~\autoref{RGS0720}. Two other absorption features at $\mathrm{0.48~keV}$ and $\mathrm{0.53~keV}$ seem even more significant with respect to the continuum (\autoref{RXJ0720_LETG}), thus, in addition to the narrow absorption feature found by \cite{2009A&A...497L...9H}, two other Gaussian absorption lines (at $\mathrm{0.48~keV}$ and $\mathrm{0.53~keV}$) are added to the fit model.\\
Fitting the data with the additional lines, the fit yields $\chi^{2}/d.o.f.=1.02$ with 249 degrees of freedom. Since all lines are rather weak, for a cross-check {\it Chandra} pipeline products were used as well as reprocessed data for the co-added spectrum together with two different absorption models ({\it phabs} and {\it tbabs}) for the ISM and different grouping of the co-added spectra using the task {\it grppha}. In all cases, the lines are present, see~\autoref{RXJ0720_LETG}. The parameters of the absorption lines are listed in~\autoref{bbfitHRC}.\\
The absorption feature at $\mathrm{0.48~keV}$ is not seen in the co-added {\it RGS} spectrum of \rxj{}, thus we also co-added all available {\it Chandra HRC-S/LETG} spectra of the much brighter NS \rxf{} (for the individual observations we refer to \citealt{2001A&A...379L..35B,2003A&A...399.1109B,2002ApJ...572..996D}) for comparison.\\
For the fitting procedure we applied the same model as in~\autoref{1856}. The fit results are listed in~\autoref{bbfitHRC} and are in good agreement to the values published in \citet{2001A&A...379L..35B,2003A&A...399.1109B}. 
The fit residuals of the {\it Chandra HRC-S/LETG} spectrum of \rxf{} do not exhibit the absorption features we found in the fit residuals of the {\it Chandra HRC-S/LETG} spectrum of \rxj{}, see~\autoref{res_0720_1856}. Thus, although weak, the features found in the spectrum of \rxj{} are not of an instrumental origin.   
\begin{figure}
\centering
\includegraphics*[viewport=105 270 490 560, width=0.425\textwidth]{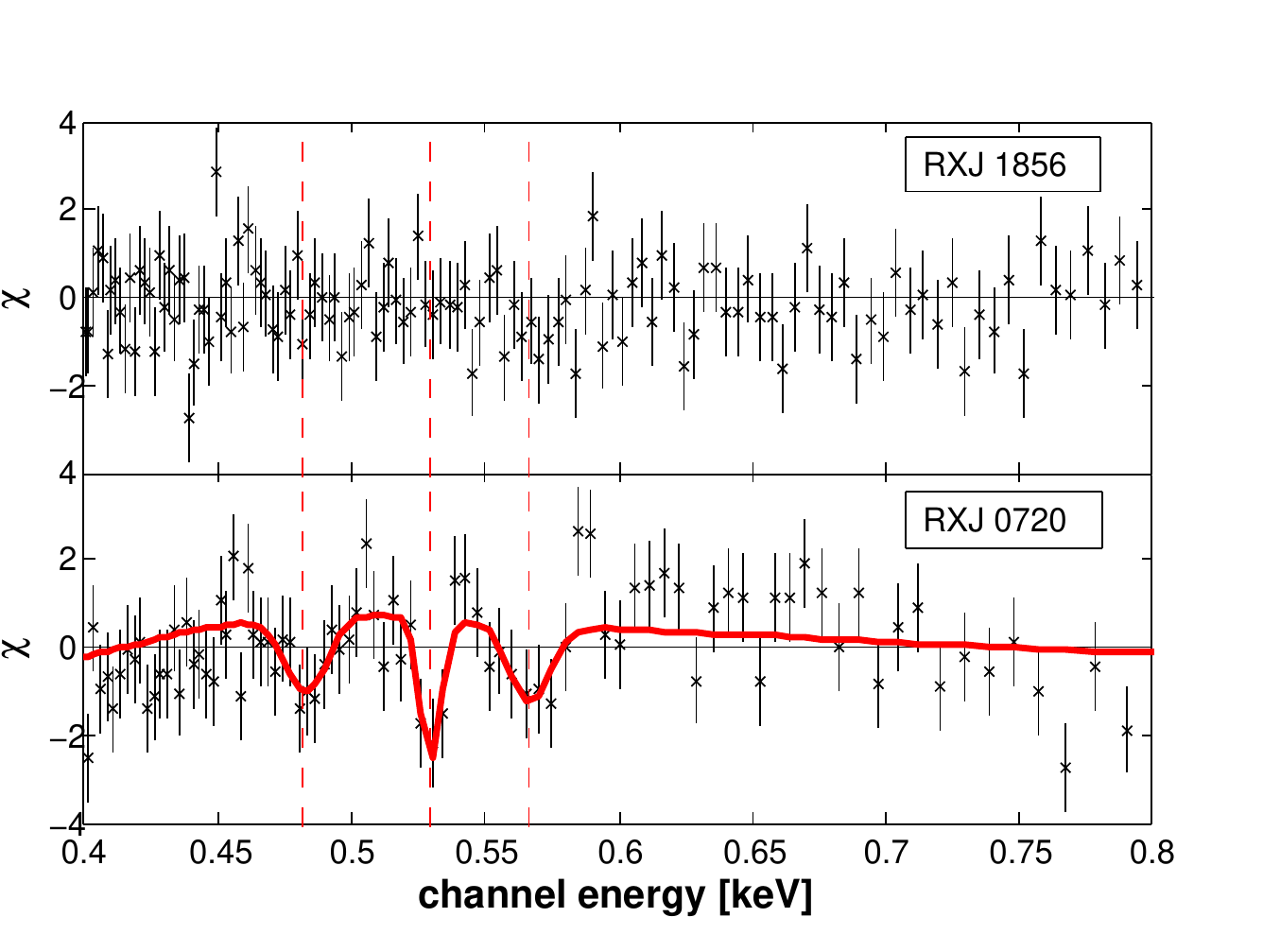}
\caption[]{The fit residuals of \rxf{} (see also \citealt{2001A&A...379L..35B,2003A&A...399.1109B}) and \rxj{} obtained from the co-added {\it Chandra HRC-S/LETG} spectra. The absorption features (red solid line in the lower panel) seen in the residuals of \rxj{} are absent in the residuals of \rxf{}. The red dashed lines mark the line centres of the three absorption features.}
\label{res_0720_1856}%
\end{figure}

\section{Discussion}
\label{conclusions}

The identification of narrow absorption lines in the X-ray spectra of highly magnetised NSs is non-trivial: If a line originates from the NS atmosphere, it is redshifted by a factor of $g_{r}=1.11$ ($M=1~\mathrm{M_{\odot}}$, $R=16~\mathrm{km}$) to $g_{r}=2.00$ ($M=2~\mathrm{M_{\odot}}$, $R=8~\mathrm{km}$) and the ion itself is not known. Considering abundances, lines in NS spectra may originate from highly ionised mid-Z elements or iron \citep{2007MNRAS.377..905M}, probably from the $K_{\alpha}$ lines, since they would cause the most strongest features. The strong magnetic field would shift and broaden the spectral lines.\\
An alternative interpretation is the interstellar origin of these absorption features, caused by the ISM, in particular in the LB, see e.g. \citet{2003A&A...411..447L} or \citet{2006A&A...452L...1B,2009SSRv..143..263B}. Some of the NSs in this work are likely located within the LB (e.g. \rxf{}), whereas other sources are outside (e.g. PSR\,B1055-52). We will discuss the two interpretations in the following sections.

\subsection{Absorption features from the neutron star atmosphere}
  
The detected line at $\mathrm{0.57~keV}$ corresponds to OVII or OVI \citep{Hirata} without gravitational redshift, as discussed in \citet{2009A&A...497L...9H}.
This line is actually a blend that mainly consists of the $K_{\alpha}$ resonance line ($\mathrm{1s^{2}-1s2p^{1}P_{1}(r)}$) at 573.95~eV of the triplet of He-like oxygen (OVII). This resonance line is superimposed with numerous weak OVI lines (see \citealt{Hirata,2009A&A...497..291K} for details), therefore the absorption feature at $\mathrm{0.57~keV}$ appears not as a perfect Gaussian and the equivalent widths and the line widths ($\sigma$) are larger than expected from the single, fully resolved lines (note, the resolution of \textit{RGS1} at 0.57~keV is $\mathrm{\approx1.7~eV}$). The intercombination line ($\mathrm{1s^{2}-1s2p^{3}P_{1,2}(i)}$) at 568.63~eV and the forbidden line ($\mathrm{1s^{2}-1s2s^{3}S_{1}(f)}$) at 560.99~eV of the OVII triplet do not contribute to the absorption feature, since they appear only in emission (not detected in our spectra).\\
If highly ionised oxygen is present close to the NS surface, the OVIII $\mathrm{Ly_{\alpha}}$ line ($\mathrm{1s-2p}$; i.e. OVIII $\mathrm{K_{\alpha}}$) at $\mathrm{653.62~eV}$ could appear at $\mathrm{\approx0.57~keV}$ with a reasonable redshift of $g_{r}=1.16$ (or $g_{r}=1.17$) and the corresponding OVII $K_{\alpha}$ resonance line should appear at $\mathrm{0.48~keV}$ ($g_{r}=1.17$). In \cite{2009A&A...497L...9H} it is argued, that such an absorption feature may be detected in the co-added {\it RGS} spectra of \rxj{}.\\
We find no evidences for such a feature, neither in the co-added {\it RGS} spectra of \rxj{}, nor in the co-added {\it RGS} spectra of \rxe{} (see~\autoref{RGS0720} or~\autoref{RGS1605}). The {\it RGS} detector exhibits an intrinsic feature at $\mathrm{0.48~keV}$ that may superimpose a weak feature and inhibit a possible detection. However, we find such an absorption feature in the co-added {\it Chandra HRC-S/LETG} spectra of \rxj{} (\autoref{RXJ0720_LETG} and \autoref{res_0720_1856}). This feature is rather weak, but comparable (regarding to its broadness) to the absorption feature at 0.57~keV (\autoref{RGS0720}), thus should be seen in the co-added {\it RGS} spectra of \rxj{} too (although superimposed by a much narrower instrumental feature).\\
Intriguingly, the absorption feature at $\mathrm{0.48~keV}$ (if real, see~\autoref{res_0720_1856}) would fit the interpretation of a gravitational redshifted line (implying a redshift of $\mathrm{g_{r}=1.17}$\footnote{With this value, \rxj{} would exceed its Schwarzschild radius by a factor of 3.7, i.e. would have an intrinsic radius of 15.6~km (assuming a mass of $\mathrm{1.4~M_{\odot}}$) or 18.3~km at infinity. However, the absorption features may originate from regions above the NS surface. Then, the given values for the radius are upper limits.}) and such features are expected at the particular energy range from NS atmosphere models (see \citealt{2007MNRAS.377..905M}, in particular Figure~14 therein for $\mathrm{10^{13}~G}$ field strength and 1~MK surface temperature, as measured for \rxj{}). On the other hand, the absorption feature at 0.48~keV is not very significant and requires further confirmation. The measured magnetic field strengths of the M7 (and the 3M) significantly increase the binding energy of atoms, if close to the NS. Under these conditions, the temperatures of $\mathrm{\approx1MK}$ might be too low to generate significant amounts of highly ionised oxygen, as discussed in \citet{2009A&A...497L...9H}. However, the atomic physics is not well understood for magnetic fields of about $\mathrm{B\approx10^{13}~G}$.\\
Note, there is no such feature (at 0.48~keV) expected from the ISM (CVI would appear in this energy range, but significantly weaker, \citealt{Hirata,2009A&A...497..291K}). An absorption feature at $\mathrm{0.65~keV}$ (OVIII at rest, see \citealt{2009A&A...497L...9H}) remains undetected, probably due to the lack of photons in this energy range (if from the ISM, the high temperatures that are required to generate OVIII are not likely).\\ 
The absorption feature found at $\mathrm{0.53~keV}$ in the {\it Chandra HRC-S/LETG} data of \rxj{} (see~\autoref{RXJ0720_LETG}) likely originates from the K-edge of neutral oxygen OI \citep{2009A&A...497..291K} at rest. This indicates an overabundance of neutral oxygen in the line of sight or might result from inexact modeling of this edge. Although we use different ISM absorption models (e.g. {\it phabs} or {\it tbabs}), this particular feature remains.\\
The absorption feature at $\mathrm{0.57~keV}$ is well detected in the \xmm{} {\it RGS} spectrum of \rxj{} \citep{2009A&A...497L...9H} and it appears (although weak) in the {\it Chandra} {\it HRC-S/LETG} spectra, i.e. its presence is certain (5.5$\sigma$ significance in the {\it RGS} spectra). Moreover, a similar feature is also detected in the co-added {\it RGS} spectra of \rxe{} (\citealt{2004ApJ...608..432V} and~\autoref{RGS1605}). The energies of the line centres are well fitted in both cases (\autoref{fit_tab}), where the line centre of the absorption feature in the spectrum of \rxe{} is shifted by $\mathrm{\approx7.0\pm4.5~eV}$ with respect to the line centre of the absorption feature in the spectrum of \rxj{}. This shift cannot be explained by the same transition of a particular ion (e.g. OVII) at rest caused by different radial motion, since the required speed would be $\mathrm{\approx3500~km/s}$. Either the two lines represent different transitions ($\mathrm{1s^{2}-1s2p^{1}P_{1}(r)}$ at 573.95~eV in the case of \rxe{} and $\mathrm{1s^{2}-1s2p^{3}P_{1,2}(i)}$ at 568.63~eV in the case of \rxj{}; both for OVII at rest) $-$ but that raises the questions why one transition is completely suppressed and why the intercombination line at $\mathrm{568.6_{-1.9}^{+1.8}~eV}$ (\rxj{}) is seen in absorption, not in emission. Alternatively, the shift might be caused by a slightly different gravitational redshift due to different NS compactness ($\mathrm{g_{r}=1.14}$ for \rxe{}) and both absorption features are caused by the OVIII $\mathrm{Ly_{\alpha}/K_{\alpha}}$ line.\\
We found a similar feature at $\mathrm{0.57~keV}$ in the co-added {\it RGS} spectrum of PSR\,B1055-52. However, the spectrum has not enough counts to identify this line undoubtedly and to determine the line energy of the line centre sufficiently accurate.\\
Note, spectral lines originating from the NS surface should appear splitted (and shifted) due to the Zeeman-effect \citep{1977ApJ...216L..67S} and broadened by pressure \citep{1997ApJ...476L..47P}, that is not considered in this interpretation. Furthermore, the magnetic field is not constant over the NS surface, i.e. the lines should be broadened due to the contributions from different surface patches.\\

\subsection{Absorption features of an interstellar or circumstellar origin}
\begin{figure*}
\centering
\includegraphics*[viewport=2 110 720 405, width=0.95\textwidth]{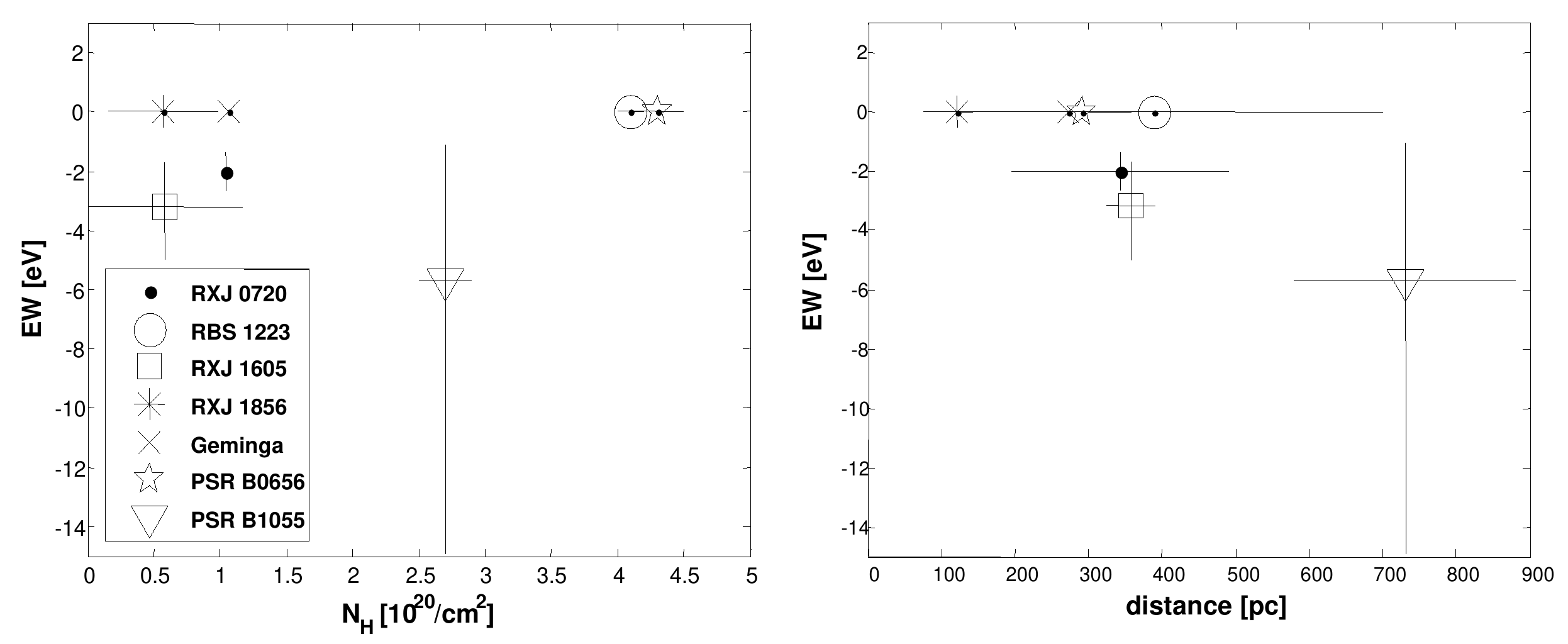}
\caption[]{Equivalent widths (EW) of the narrow absorption features at 0.57~keV (errors denote 90\% confidence) as measured in this work compared to $\mathrm{N_{H}}$ and the distances of the ``Magnificent Seven" and the ``Three Musketeers". The values and errors for $\mathrm{N_{H}}$ are obtained from {\it EPIC-pn} spectra \citep{2009A&A...498..811H,2003A&A...403L..19H,2004ApJ...608..432V}, in the case of the ``Three Musketeers" also from both {\it EPIC-MOS} \citep{2005ApJ...623.1051D} and in the case of \rxf{} from {\it EPIC-pn} and {\it Chandra HRC-S/LETG} spectra \citep[note the different values for $\mathrm{N_{H}}$ obtained for {\it EPIC-pn} and {\it LETG} as published therein]{2001A&A...379L..35B,2003A&A...399.1109B}, since these spectra cover more of the soft part of the emission of the neutron stars than {\it RGS}, thus yield more reliable values for $\mathrm{N_{H}}$. The individuell distances are obtained from the literature cited in the sections 1.1, 1.2. and 2.}
\label{dist_EW_NH}%
\end{figure*}
The distances of \rxj{} and \rxe{} are $\mathrm{\approx195-530~pc}$ (from direct parallax measurements, \citealt{2007ApJ...660.1428K,EisenPhd}) and $\mathrm{\approx325-390~pc}$ (estimated from extinction models, \citealt{2007Ap&SS.308..171P}), respectively. Given the average volume density of OVII $\mathrm{n_{OVII}=(1.35-2.84)\times10^{-6}/cm^{3}}$ in the ISM \citep[assuming that oxygen is completely ionised to OVII]{2005ApJ...624..751Y}, the expected OVII column densities for these two NSs are $\mathrm{N_{OVII}=1.2\times10^{14}-4.1\times10^{15}/cm^{2}}$ and a factor of two or three larger for PSR\,B1055-52, since this NS is more distant \citep{2003MNRAS.342.1299K}. The hydrogen column densities from these three NSs are in the order of $\mathrm{N_{H}=10^{20}/cm^{2}}$ (\autoref{fit_tab}). Applying the oxygen abundance relative to hydrogen ($\mathrm{n_{O}/n_{H}=4\times10^{-4}}$, \citealt{1989GeCoA..53..197A}), the expected column density of OVII should be in the order of $\mathrm{N_{OVII}=10^{16}/cm^{2}}$ (again assuming that oxygen is completely ionised to OVII).\\ 
The measured equivalent widths of the narrow absorption features at 0.57~keV yield $\mathrm{N_{OVII}=3\times10^{16}-10^{19}/cm^{2}}$ (\rxj{}), $\mathrm{N_{OVII}=3\times10^{16}-10^{20}/cm^{2}}$ (\rxe{}) and $\mathrm{N_{OVII}\geq10^{16}/cm^{2}}$ (PSR\,B1055-52), taking fit errors (\autoref{fit_tab}) and different velocity dispersions \citep{2004ApJ...605..793F} into account and assuming that the absorption feature at 0.57~keV is a pure line (not a blend).\\
\citet{2009A&A...497L...9H} discussed a possible circumstellar origin of the absorption feature at 0.57~keV in the \textit{RGS} spectra of \rxj{} to explain this difference of the column densities and showed that the absorption feature might be (partly) caused by an ambient medium surrounding \rxj{} in $\mathrm{\approx10^{5}~km}$ distance from the NS, superimposed by absorption lines from the ISM. The model in \citet{2009A&A...497L...9H} is also applicable to \rxe{} and PSR\,B1055-52, since both NSs have temperatures and magnetic field strengths comparable to that of \rxj{}. However, circumstellar disks surrounding isolated NSs with ages of the order of million years should be rare. Thus, it is unlikely that three sources out of seven in our sample host a disk.\\
Regarding the equivalent widths and the significance of the detection, the strongest narrow absorption features are present in the spectra of \rxj{} and \rxe{}. Both NSs are located outside the LB \citep{2003A&A...411..447L}, i.e. are embedded in a denser medium than in the solar vicinity. The LB hosts hot plasma containing highly ionised oxygen \citep{2006A&A...452L...1B,2009SSRv..143..263B}, such as OVII or OVI that is in the line of sight to these NSs. In contrast, the closest NS, \rxf{}, is almost certainly located within the LB (the most recent parallax measurement yields a distance of $\mathrm{\approx125~pc}$, see \citealt{2010arXiv1008.1709W}) towards the Galactic centre, i.e. embedded in a thin medium and therefore should not exhibit the narrow absorption feature.\\
The distance of \rxd{} (\rbs{}) is highly uncertain and ranges from $\mathrm{\approx76-700~pc}$ (see \citealt{2007Ap&SS.308..181H} and references therein), but this NS has a high Galactic latitude and is located either within the LB (for smaller distances), or above the Galactic plane. Even in the latter case, \rxd{} is still surrounded by a low density medium (the local chimney, see \citealt{2003A&A...411..447L} who shows the distribution of measured column densities). Therefore, also the spectra of \rxd{} should not exhibit narrow absorption features.\\
Geminga and PSR\,B0656+14 are located at roughly the same region above the Galactic plane, probably within the local chimney (the distances range from 150~pc to 300~pc for both objects, see \citealt{1996ApJ...461L..91C,2003ApJ...593L..89B,2005nscf.confE..23W}), like in the case of \rxd{}. Both NSs do not exhibit narrow absorption features, although in the case of Geminga the count number of the co-added {\it RGS} spectrum is too low to allow a final statement.\\     
Finally, the spectrum of PSR\,B1055-52 shows an uncertain absorption feature and this NS is likely located outside the LB and is probably the most distant object in our sample (\citealt{2003MNRAS.342.1299K} estimated 750~pc from dispersion measurements).\\
If caused by the ISM, other X-ray sources (particularly much more distant) like e.g. Cyg~X$-$2~\citep{2009ApJ...696.1418Y}, LMXB~GS~1826$-$238~\citep[and other Galactic low-mass X-ray binaries; \citealt{2005ApJ...624..751Y}]{2010A&A...521A..79P} or Mrk~421~\citep[with 955~ks exposure time]{2007ApJ...656..129R,2006ApJ...652..189K} should exhibit similar absorption features. While the expected OVII resonance line at 573.95~eV of an interstellar origin is clearly visible (in particular in \citealt{2006ApJ...652..189K}, Figure 2 therein and \citealt{2007ApJ...656..129R}, Figure 1 therein), these sources do not show an absorption feature at 568.6~eV in their \textit{RGS} spectra as detected in the case of \rxj{}. The line energy in the {\it Chandra} {\it HRC-S/LETG} spectra of \rxj{} yields $\mathrm{566.4_{-7.2}^{+6.6}~eV}$ (\autoref{bbfitHRC}), i.e. this value is almost compatible (within its errors) with the energy of the OVII $\mathrm{K_{\alpha}}$ resonance line at rest.

\section{Conclusions}

We investigated the co-added {\it RGS} spectra of the four brightest M7 (\rxj{}, \rxd{}/\rbs{}, \rxe{} and \rxf{}) and the 3M; Geminga, PSR\,B0656+14 and PSR\,B1055-52 and searched for narrow absorption features, in particular at 0.57~keV.\\ 
We find that those NSs, that are either nearby or located in a medium with low densities do not exhibit narrow absorption features, whereas those NSs, that are distant and surrounded by a dense medium do exhibit narrow absorption features. The equivalent widths of the narrow absorption features at 0.57~keV do not correlate with the $\mathrm{N_{H}}$ value, and hardly correlate with the distances (\autoref{dist_EW_NH}).\\
Other well investigated X-ray sources (Cyg~X$-$2 \citealt{2009ApJ...696.1418Y}, LMXB~GS~1826$-$238 \citealt{2010A&A...521A..79P}, Mrk~421 \citealt{2007ApJ...656..129R,2006ApJ...652..189K}) do not exhibit an absorption feature at 568.6~eV as detected in the case of e.g. \rxj{} (\textit{RGS}). If this feature would have an origin in the NS atmosphere, the gravitational redshift and, thus, the compactness points either to a large radius and/or a small mass. However, we stress that the underlying atomic physics is not yet well understood for high magnetic field strengths.

\section*{Acknowledgments}

The \xmm\ project is supported by the Bundesministerium f\"ur Wirtschaft und
Technologie/Deutsches Zentrum f\"ur Luft- und Raumfahrt (BMWI/DLR, FKZ 50 OX 0001)
and the Max-Planck Society.
MMH acknowledges support by the Deutsche Forschungsgemeinschaft (DFG) through
SFB/TR 7 ``Gravitationswellenastronomie''.
We thank Valeri Hambaryan for fruitfull discussions, who encouraged us for this work, and the anonymous referee for helpfull comments.

\bibliography{references}


\end{document}

%% file: definitions.tex
\newcommand{\xmm}{{\it XMM-Newton}}

\newcommand{\rxj}{\hbox{{RX\,J0720.4$-$3125}}}

\newcommand{\rxd}{\hbox{{RX\,J1308.6+2127}}}
\newcommand{\rxe}{\hbox{{RX\,J1605.3+3249}}}
\newcommand{\rxf}{\hbox{{RX\,J1856.4$-$3754}}}
\newcommand{\rbs}{\hbox{{RBS1223}}}

\def\gtrsim{\mathrel{\hbox{\rlap{\hbox{\lower4pt\hbox{$\sim$}}}\hbox{$>$}}}}